\documentclass[11pt,a4paper,twoside]{article}

\usepackage[utf8]{inputenc}	
\usepackage{amssymb}
\usepackage{amsmath}
\usepackage{bm}
\usepackage{graphicx}
\usepackage{subcaption}
\usepackage[title]{appendix}

\usepackage[affil-it,auth-lg]{authblk}
\usepackage{tabularx}
\usepackage{booktabs}
\usepackage{diagbox}
\usepackage[mathscr]{euscript}

\usepackage[dvips]{color} % dvips allows for colors

\allowdisplaybreaks

\begin{document}
%	\linenumbers
	
\title{Detectable universes inside regular black holes}
\author[1,2]{Zacharias Roupas}
\affil[1]{Department of Physics, University of Crete, 70013, Herakleion, Greece} 
\affil[2]{The British University in Egypt, Sherouk City 11837, Cairo, Egypt}

\date{\vspace{-5ex}}

%\pacs{}
%\keywords{} 	

\maketitle

	%\pacs{}
	%\keywords{} 	
	
\begin{abstract}
	While spacetime in the vicinity outside astrophysical black holes is believed to be well understood, the event horizon and the interior remain elusive. Here, we discover a degenerate infinite spectrum of novel general relativity solutions with the same mass-energy and entropy that describe a dark energy universe inside an astrophysical black hole. This regular cosmological black hole is stabilized by a finite tangential pressure applied on the dual cosmological-black hole event horizon, localized up to a quantum indeterminacy. We recover the Bekenstein-Hawking entropy formula from the classical fluid entropy, calculated at a Tolman temperature equal to the cosmological horizon temperature. We further calculate its gravitational quasi-normal modes. We find that cosmological black holes are detectable by gravitational-wave experiments operating within the $\mu{\rm Hz}-{\rm Hz}$ range, like LISA space-interferometer.  
\end{abstract}

\section{Introduction}

	As early as 1966, Sakharov \cite{1966JETP...22..241S} proposed that the proper equation of state of matter and energy at very high densities is that of a dark energy fluid $P = -\rho c^2$. About the same time Gliner \cite{1966JETP...22..378G} suggested that a spacetime filled with vacuum could provide a proper description of the final stage of gravitational collapse, replacing the future singularity \cite{1966JETP...22..378G}. Black hole solutions where the singularity is avoided are called regular black holes \cite{1968_Bardeen,BRONNIKOV197984,1981NCimL......161G,1988CQGra...5L.201P,1992GReGr..24..235D,2013PhLB..721..329B,2018JCAP...02..059R,Ansoldi:2008jw} and may or may not involve a de Sitter core. The, so called, dark energy stars or gravastars \cite{2001gr.qc.....9035M,2004CQGra..21.1135V,2005CQGra..22.4189C,2006CQGra..23.1525L,2007CQGra..24.4191C,2019PhRvD..99h4021B,2019PhRvD..99d4037B,2020IJMPD..2930004R,2020MPLA...3550071B} generally do not predict the presence of an event horizon. 
	
	The idea that a new universe is generated inside a black hole has been put forward in \cite{1972Natur.240..298P,1989PhLB..216..272F,1990PhRvD..41..383F,Poplawski_2016,Smolin_1997}.
	Gonzalez-Diaz \cite{1981NCimL......161G} was, to our knowledge, the first to explicitly propose that a de Sitter space may complete an exterior Schwartzschild metric with the presence of a kind of cosmological black hole horizon in-between. Later, it was realized by Poisson \& Israel that in this case a singular tangential pressure will be exerted on the horizon \cite{1988CQGra...5L.201P}.
The present work elaborates on the Poisson-Israel solution regularizing the horizon by considering that the quantum uncertainty principle applies.

We discover an infinite spectrum of solutions, which describe the fluid shell which matches the interior de Sitter core with the exterior Schwarzschild spacetime. The metric's derivatives are continuous up to any required order. All states of the spectrum have the same energy and entropy. This fluid entropy of the dual horizon recovers the Bekenstein-Hawking black hole entropy for a fluid temperature equal to the cosmological horizon temperature. This spacetime spectrum describes a novel kind of regular black hole we shall call the ``cosmological black hole'' for brevity.

Gravitational-wave astronomy has opened up the possibility to detect such objects. The cosmological black holes may exist independently than singular or other types of regular black holes, or may describe the state of all detected black holes. The detectability of cosmological black holes is founded on the fact that the fundamental quasi-normal mode, calculated here, is distinctively different than the one of Schwarzschild black holes for any mass. These modes are closely related to the ringdown phase of a post-merger object. This phase follows the inspiral phase of a binary black hole merger. The ringdown phase is dominated by the natural frequencies of black hole spacetime, like a ringing bell. We argue that LIGO-Virgos's detections could involve cosmological black holes, because LIGO-Virgo is not able to discriminate between cosmological and singular black holes, due to the well-known ``mode camouflage'' mechanism \cite{2014PhRvD..90d4069C} and the inadequate frequency sensitivity. On the other hand, the frequency spectrum of quasi-normal modes of the cosmological black hole interior lie within the detectability range frequencies of the planned LISA space interferometer.

\section{The cosmological black hole solution spectrum}\label{sec:DE-BH}

A black hole is formed from material that crossed the Schwartzschild horizon. Thus, inside the horizon it is proper to use the full Einstein equations $G_{\mu\nu} = (8\pi G/c^4) T_{\mu\nu} $ instead of the field vacuum equation $G_{\mu\nu} = 0$. Outside the horizon the Schwartzschild metric should apply assuming that all of material has crossed the horizon. One such solution requires the interior to be a de Sitter vacuum $P = -\rho_0 c^2 = {\rm const.}$, with a tangential pressure $P_{\rm T} = -\rho_0 c^2\theta\left(1-\frac{r}{r_{\rm H}} \right) + \frac{1}{2} \rho_0  c^2\delta \left(\frac{r}{r_{\rm H}} - 1 \right)$ being applied on the horizon $r_{\rm H}$, where $\theta$ and $\delta$ denote the Heaviside and Dirac functions respectively. This expression was mentioned (without a derivation) for the first time, to our knowledge, by Poisson \& Israel  \cite{1988CQGra...5L.201P}. We derive in detail this solution in Appendix \ref{app:P-I}. Poisson \& Israel remarked that this tangential pressure diverges for an observer at some proper distance outside the horizon (see equation (\ref{eq:P_T_divergence}) of Appendix \ref{app:P-I}). Nevertheless, we see little qualitative difference regarding the physical problems encountered by the Poisson-Israel solution and the Schwarzschild black hole solution, which does present a curvature singularity in the centre. It is only that the problem in the former case is transferred from the center to the horizon, having the curvature singularity of the Schwarzschild solution replaced by a pressure singularity in the Poisson-Israel solution.

However, assuming that within $r_{\rm H}$ there is distributed a mass $M_\bullet$ in some non-singular way up to $r_{\rm H}$, quantum physics suggests that the boundary $r_{\rm H}$ of $M_\bullet$ cannot be localized with accuracy greater than the Compton wavelength  
\begin{equation}
	\Delta r_{\rm H} \gtrsim \frac{h}{M_\bullet c}.
\end{equation}
It is justified therefore to assume there exists a length-scale $\alpha$ that specifies the quantum fuzziness of the horizon 
\begin{equation}\label{eq:Delta_rH}
	\Delta r_{\rm H}=\alpha 
\end{equation}
in this case.
For an astrophysical black hole with mass $M_\bullet =  \mathscr{O}({\rm M}_\odot)$, $\alpha$ may equal the Compton wavelength or the Planck scale or a few times the latter, so that 
\begin{equation}\label{eq:varepsilon}
	\varepsilon \equiv \frac{\alpha}{r_{\rm H}} \ll 1
\end{equation}
in each of these cases. Lacking a quantum theory of gravity we cannot know its precise value, still we shall be able here to reach definite quantitative results, irrespective from the value of $\alpha$. As we shall now show, the Poisson-Israel solution gets regularized by an infinite spectrum of solutions with the same energy and entropy.

\begin{figure}[!tb]
	\begin{center}
		\begin{subfigure}{0.45\textwidth}
			\includegraphics[scale = 0.5]{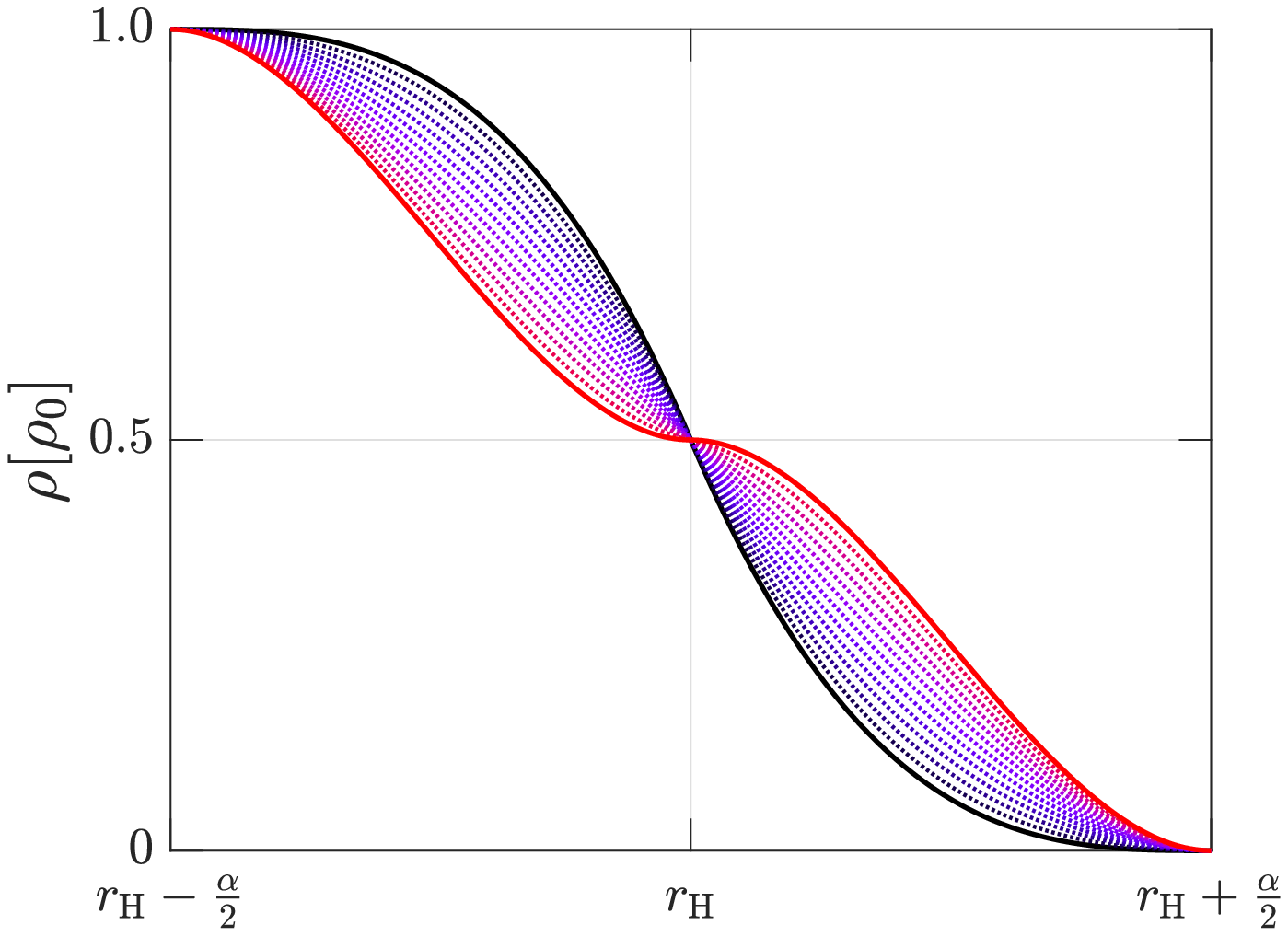}  
			\caption{Density ($K=1$, $N=3$).}
			\label{fig:rho_1st_order_spectrum}
		\end{subfigure}
		\begin{subfigure}{0.45\textwidth}
			\includegraphics[scale = 0.5]{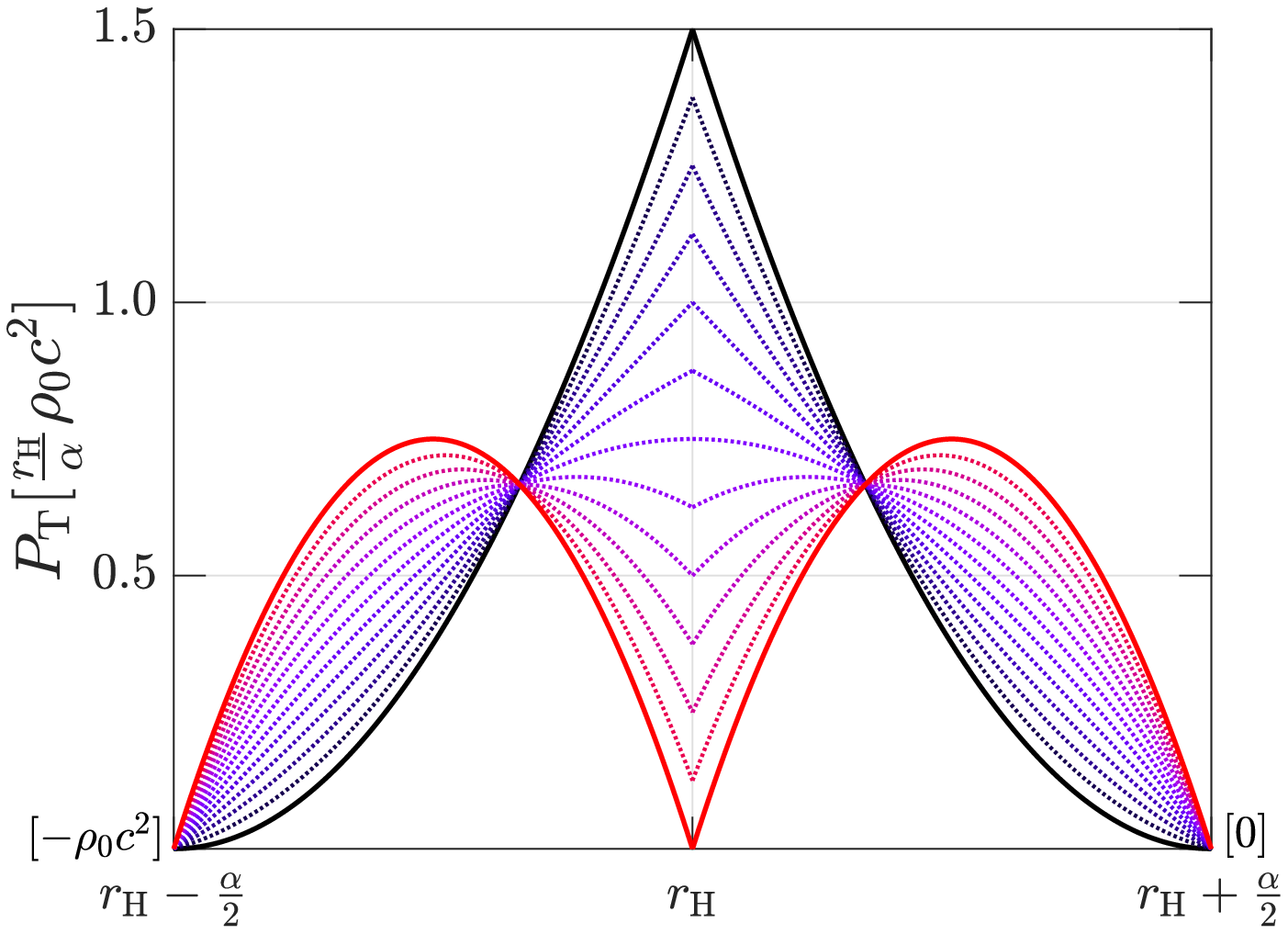}  
			\caption{Tangent pressure ($K=1$, $N=3$).}
			\label{fig:P_T_1st_order_spectrum}
		\end{subfigure}
		\\
		\begin{subfigure}{0.45\textwidth}
			\includegraphics[scale = 0.5]{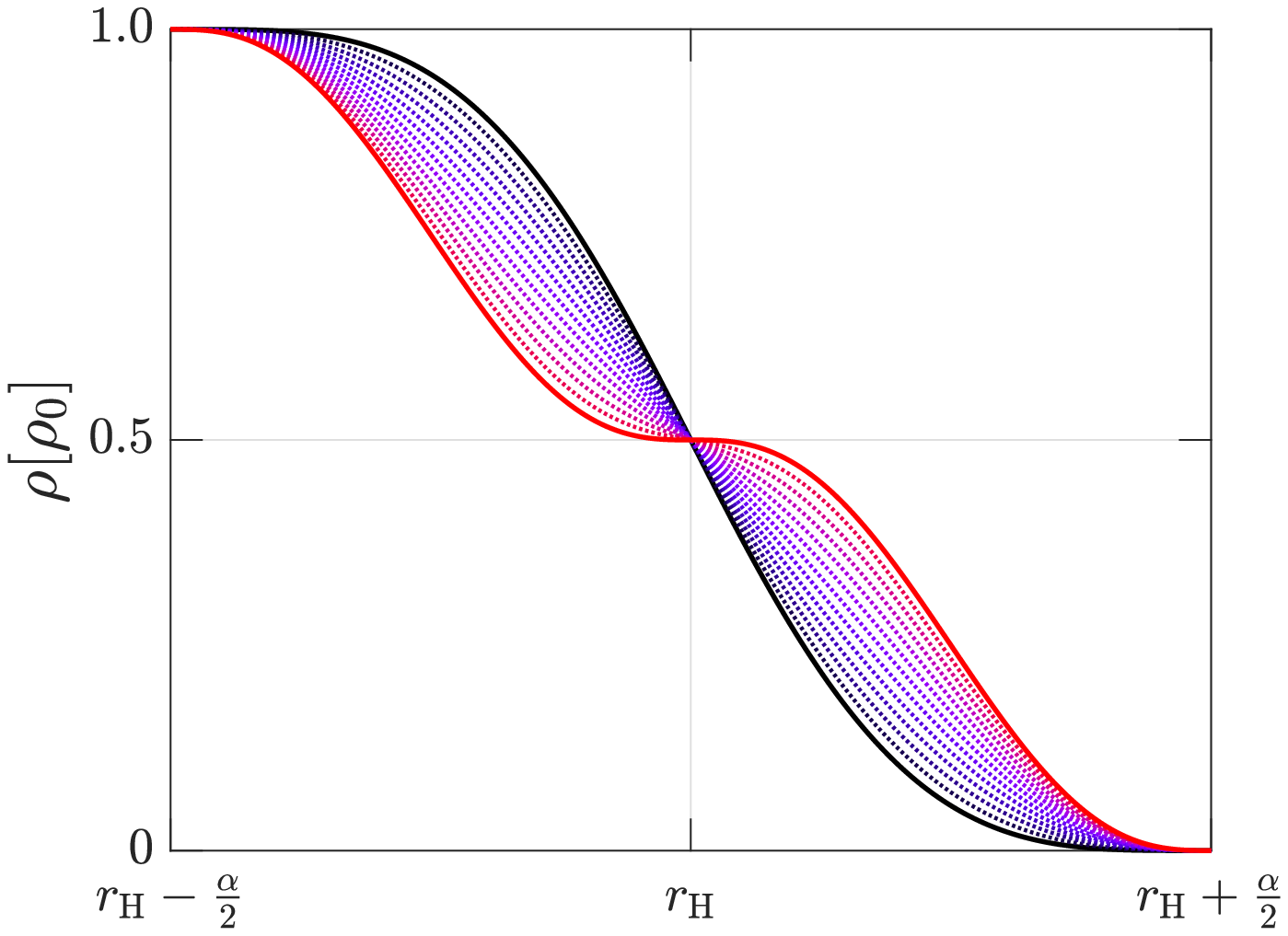}  
			\caption{Density ($K=2$, $N=5$).}
			\label{fig:rho_2nd_order_spectrum}
		\end{subfigure}
		\begin{subfigure}{0.45\textwidth}
			\includegraphics[scale = 0.5]{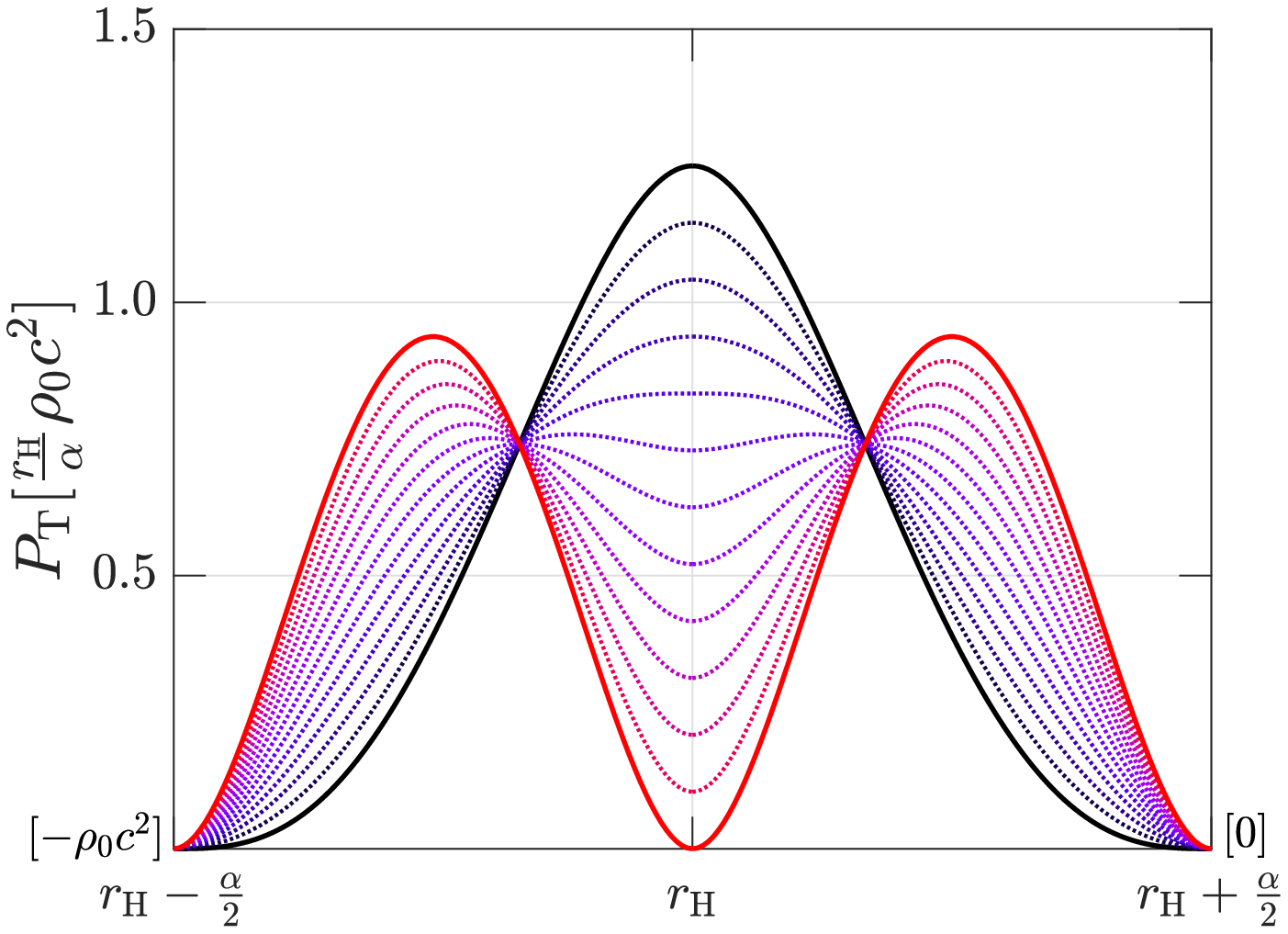}  
			\caption{Tangent pressure ($K=2$, $N=5$).}
			\label{fig:P_T_2nd_order_spectrum}
		\end{subfigure}
	\captionsetup{justification=raggedright}
		\caption{The cosmological black hole spectrum of density (\ref{eq:solutions})  and tangential pressure (\ref{eq:P_T}) inside the $\alpha$-shell, that designates the horizon. The values of $P_{\rm T}$ on the boundaries are mentioned separately in brackets, because the scaling does not allow them to be distinguished optically. The value of the radial pressure is $P_{\rm r}(r) = -\rho(r)c^2$ everywhere. Exterior to the horizon, $\rho$, $P_{\rm T}$ are zero and in the interior $P_{\rm T} = -\rho c^2 = -\rho_0 c^2 = {\rm const}$. Each curve in all panels represents a solution of the Einstein equations with the same energy and entropy. The solid curves represent the limiting allowed curves for which $\rho^\prime \leq 0$ everywhere, for the respective maximum order of continuous metric derivatives. \textit{Upper panels}: The metric, its first and second derivatives are continuous everywhere, that is $K=1$. We considered $N=3$ and each curve corresponds to a different value of $A_2^{(3)} \in {[-4,8]}$. \textit{Lower panels}: The metric, its first, second and third derivatives are continuous everywhere, that is $K=2$. We considered $N=5$ and each curve corresponds to a different value of  $A_1^{(5)} = A_2^{(5)} \in {[-96,24]}$.}
		\label{fig:spectrum}
	\end{center}
\end{figure}

Let us assume the static, spherically symmetric ansatz for the metric
\begin{equation}\label{eq:ds2_ansatz}
	ds^2 = - h(r) c^2 dt^2 + h(r)^{-1} dr^2 + r^2 d\Omega
\end{equation}
and  the following components of an anisotropic diagonal energy-momentum tensor $T_0^{~0} = -\rho(r) c^2$, $T_1^{~1} = P_{\rm r} (r)$, $T_2^{~2}=T_3^{~3} = P_{\rm T}(r)$, where $\rho(r)$ is the mass density and $P_{\rm r}(r)$, $P_{\rm T}(r)$ the radial and tangential pressures.  
It is straightforward to show (see Appendix \ref{app:P-I}) that the Einstein equations admit the following formulation
\begin{equation}
	\label{eq:dmdr}
	h(r) = 1 - \frac{2Gm(r)}{r c^2},
	\quad 
	\frac{d m(r)}{dr} = 4\pi \rho(r) r^2,
\end{equation}
and
\begin{align}
\label{eq:P_r}	P_{\rm r}(r) &= - \rho(r) c^2 \\
\label{eq:P_T}	P_{\rm T}(r) &= - \rho(r) c^2 - \frac{1}{2}r\rho(r)^\prime c^2.
\end{align}
Note that another density distribution function, besides the Poisson-Israel solution (\ref{eq:m_PI})-(\ref{eq:P_T_PI}), that solves this system was identified in Ref. \cite{1992GReGr..24..235D}.

We discover here a new infinite spectrum of solutions that regularizes the Poisson-Israel solution within the fuzziness $\alpha$ of the horizon
	\begin{equation} 
		\label{eq:solutions}
		\rho(r) = \left\lbrace
		\begin{array}{ll}
			\rho_0 &, \; r \leq r_{\rm H} - \frac{\alpha}{2}, \\
			\rho_{(-)}(r) &, \;  r_{\rm H} - \frac{\alpha}{2} \leq r \leq r_{\rm H}, \\
			\rho_{(+)}(r) &, \;  r_{\rm H} \leq r \leq r_{\rm H} + \frac{\alpha}{2} , \\
			0 &, \; r \geq r_{\rm H} + \frac{\alpha}{2}. \\
		\end{array}
		\right. 
,
\quad	
\rho_{(\pm)}(r) = \rho_0\sum_{n=0}^N A^{(\pm)}_n (\varepsilon) \, x(r)^n ,
\end{equation} 
where
\begin{equation}\label{eq:x_nond}
	 x \equiv \frac{r - r_{\rm H}}{\alpha} \in {[-\frac{1}{2},+\frac{1}{2}]}
\end{equation}
and $\varepsilon$ is given in (\ref{eq:varepsilon}).
Proper choices of $A^{(\pm)}_n(\varepsilon)$ ensure that the density and consequently the metric through (\ref{eq:dmdr}) are continuous and have continuous derivatives. The maximum order of the continuous derivatives can be arbitrarily high.
This is ensured by demanding to hold the following conditions
\begin{align}
\label{eq:cond_1}	
&\rho_{(-)}(r_{\rm H}-\frac{\alpha}{2}) = 	\rho_0 , 	\quad
\rho_{(-)}(r_{\rm H}) = \rho_{(+)}(r_{\rm H}) , 
\quad 
\rho_{(+)}(r_{\rm H}+\frac{\alpha}{2}) = 	0 , 	\\
\label{eq:cond_2}	
	&\frac{d^{(k)}\rho_{(-)} (r_{\rm H}-\frac{\alpha}{2})}{dr^k} = 	0 ,
\;
\frac{d^{(k)}\rho_{(-)} (r_{\rm H})}{dr^k} = \frac{d^{(k)}\rho_{(+)} (r_{\rm H})}{dr^k}, 	
\; 
\frac{d^{(k)}\rho_{(+)} (r_{\rm H}+\frac{\alpha}{2})}{dr^k}  = 0 , 
\quad 
k = 1, 2, \ldots, K
\\
\label{eq:cond_3}	
&\int_{r_{\rm H}-\frac{\alpha}{2}}^{r_{\rm H}} 4\pi \rho_{(-)}r^2 dr + \int_{r_{\rm H}}^{r_{\rm H}+\frac{\alpha}{2}} 4\pi \rho_{(+)}r^2 dr 	= \frac{4}{3}\pi \rho_0\left( r_{\rm H}^3 - \left( r_{\rm H} - \frac{\alpha}{2} \right)^3 \right) .
\end{align}
The condition $K\geq 1$ ensures that the metric and its first and second derivatives are continuous, (over)satisfying Lichatowich junction conditions, as well as that the tangential pressure (\ref{eq:P_T}) is continuous. The condition (\ref{eq:cond_3}) certifies that the total mass of the system up to the radius $R=r_{\rm H} + \frac{\alpha}{2}$ is equal to 
\begin{equation}
	M_\bullet = \frac{4}{3} \pi \rho_0 r_{\rm H}^3
\end{equation}
and therefore that  
\begin{equation}\label{eq:r_H}
	r_{\rm H} \equiv \frac{2GM_\bullet}{c^2} = \sqrt{\frac{3c^2}{8\pi G \rho_0}} .
\end{equation}
This means that at $r=r_{\rm H}$ coincide a cosmological and a black hole event horizon if the quantum indeterminacy is $\Delta r_{\rm H} = \alpha$. This renders our solution a regular, free of singularities, type of black hole, which we call the cosmological black hole solution. In the interior the scalar curvature $\mathcal{R}=6/r_H^2 \propto M_\bullet^{-2}$ is finite. We shall see in the next section that the entropy of the cosmological black hole equals the Bekenstein-Hawking entropy at the cosmological horizon temperature. Note that expressing $\rho_0$ with respect to the horizon radius
\begin{equation}
	\rho_0 = \frac{3c^2}{8\pi G}\frac{1}{r_{\rm H}^2} ,
\end{equation}
we get precisely the definition of the critical energy density in a Friedmann cosmological model for $r_{\rm H} = c/H_0$, where $H_0$ denotes the Hubble parameter.

\begin{figure}[!tb]
	\begin{center}
		\begin{subfigure}{0.45\textwidth}
			\includegraphics[scale = 0.5]{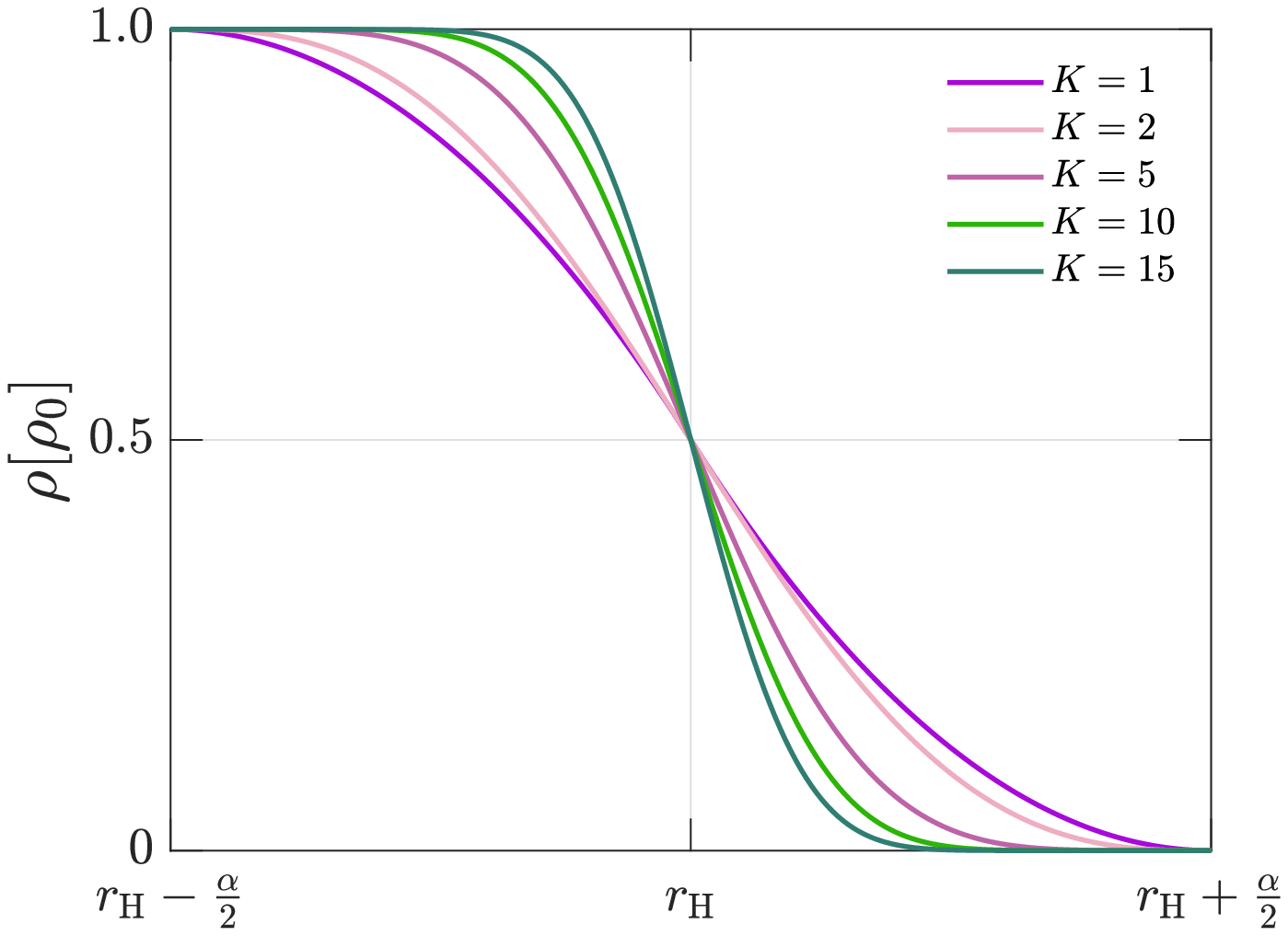}  
			\caption{Density (minimum polynomials order).}
			\label{fig:rho_orders_spectrum}
		\end{subfigure}
		\begin{subfigure}{0.45\textwidth}
			\includegraphics[scale = 0.5]{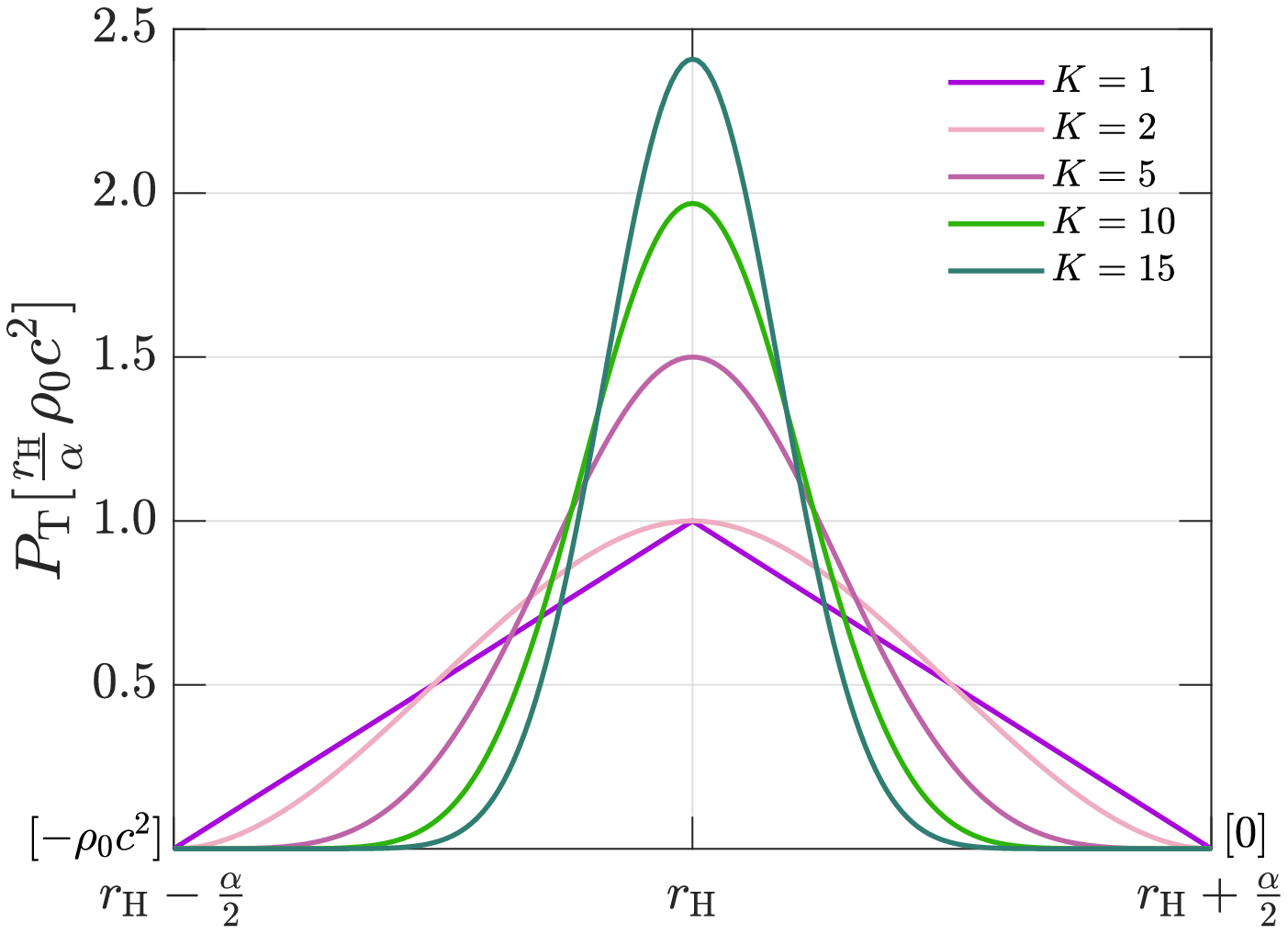}  
			\caption{Tangent pressure (minimum polynomials order).}
			\label{fig:P_T_orders}
		\end{subfigure}
	\captionsetup{justification=raggedright}
		\caption{The axis are as in Figure \ref{fig:spectrum}, where here is shown the solution with minimum order $N$ of each density polynomial (\ref{eq:solutions}) for various values of $K$ which is the maximum order of the derivative of density that is continuous. We observe that density variation is constrained for large $K$ within much smaller region than $\alpha$.}
		\label{fig:orders}
	\end{center}
\end{figure}

The order $N$ of the density polynomial (\ref{eq:solutions}) can be arbitrarily high independently from the order $K$ (\ref{eq:cond_2}) that designates the maximum order of continuous density derivatives. There can always be found solutions as long as $N \geq (3K+2)/2$. The maximum order of continuous metric derivatives is $K+1$. In the Appendix \ref{app:DE-BH} we provide the exact $K=1$ spectrum for $N=3$ and the exact $K=2$ spectrum for $N=5$ requiring that density is a decreasing function of radius
\begin{equation}
	\rho^\prime(r) \leq 0 .
\end{equation} 
This condition, along with conditions (\ref{eq:cond_1})-(\ref{eq:cond_3}), certify that the equation of state (\ref{eq:P_r}), (\ref{eq:P_T}) for the solution (\ref{eq:solutions}) constrained by these conditions satisfies the Weak Energy Condition $T_{\mu\nu}\xi^\mu\xi^\nu \geq 0$ for any time-like $\xi^\mu$; namely that $\rho^\prime \leq 0$, $P_r + \rho \geq 0$, $P_T+\rho\geq 0$.

We plot the spectrum $K=1$ for $N=3$ and the spectrum $K=2$ for $N=5$ in Figure \ref{fig:spectrum}. Note that there exist also solutions, that are not symmetrical about the $r=r_{\rm H}$ vertical axis. Furthermore, we remark that the fuzziness $\alpha$ does not necessirily constrain maximally the density variations. For large $K$, the density variation is localized within a region smaller than $\alpha$. This is depicted in Figure \ref{fig:orders}.

\section{Fluid entropy}\label{app:entropy}

The work performed by a fluid with stress tensor $T^{ij}$, $i,j=1,2,3$ may be written with respect to the strain tensor $\sigma ^i_j$ as 
\begin{equation}
	dW = T^j_i d\sigma^i_j .	
\end{equation}
The strain tensor can be decomposed as the sum of a pure shear (shape deformations) and a hydrostatic compression (volume deformations) \cite{landau1986theory}\footnote{Page 10.}
\begin{equation}
	\sigma ^i_j = \underbrace{\left(\sigma^i_j - \frac{1}{3}\delta^i_j {\rm Tr}(\sigma^i_j) \right)}_{{\rm pure}\; {\rm shear}} + \underbrace{\left(\frac{1}{3}\delta^i_j {\rm Tr}(\sigma^i_j)\right)}_{{\rm hydrostatic} \; {\rm compression}}.
\end{equation}
The trace of strain expresses relative volume change, so that considering unit volume deformations we get in general $dV = d{\rm Tr}(\sigma^i_j)$ \cite{landau1986theory}. For a spherical anisotropic fluid the space component of the energy-momentum tensor may be written in spherical coordinates as $T^i_j = {\rm diag}{(P_{\rm r},P_{\rm T},P_{\rm T})}$. The work of the gravitational force under a spherical deformation (no pure shear) is therefore
\begin{equation}
	dW = P dV, \quad {\rm where}\; P = \frac{1}{3}(P_{\rm r} + P_{\rm T} + P_{\rm T}) = P_{\rm r} + \frac{2}{3}(P_{\rm T} - P_{\rm r}).
\end{equation}
It is $P$ that contributes to the work and not only $P_{\rm r}$ despite the deformation being isotropic. There is an additional, to $P_{\rm r}$, contribution coming from $\frac{2}{3}(P_{\rm T} - P_{\rm r})$ due to the stretching forces on the fluid sphere during any spherical deformation (during a spherical expansion/contraction the area of the sphere increases/decreases, therefore there are applied tangential forces).
The relativistic thermodynamic Euler relation should involve the pressure that contributes to the work. Thus, for zero chemical potential, inside the $\alpha$-shell the thermodynamic Euler relation is properly written as
\begin{equation}\label{eq:Euler}
	Ts = \rho c^2 + P \Rightarrow s = -\frac{c^2}{3T} r\rho^\prime,
\end{equation} 
where $s=s(r)$ is the total entropy density, including the tangential contribution. We denote $T=T(r)$ the local temperature and we have used equation (\ref{eq:P_T}). Local temperature obeys the Tolman law
\begin{equation}\label{eq:Tolman}
	T(r)\sqrt{g_{tt}(r)} = T_0 = {\rm const.}
\end{equation}
where the constant $T_0$ is called the Tolman temperature and corresponds to the temperature of the fluid measured by an oberver at infinity.
The interior, excluding the horizon, does not contribute to the fluid entropy since $Ts_{\rm interior} = \rho_0 c^2 + P_{\rm interior} = \rho_0 c^2 - \rho_0 c^2 = 0$.
Thus, the total fluid entropy (integrating the local entropy over the proper volume in General Relativity) equals the fluid entropy of the event horizon, which using (\ref{eq:Euler}), (\ref{eq:Tolman}), is equal to
\begin{equation}
	S = \int_{r_{\rm H}-\frac{\alpha}{2}}^{r_{\rm H} + \frac{\alpha}{2}} s(r) \sqrt{g_{rr}}4\pi r^2dr = 
	- \frac{c^2}{3T_0}\int_{r_{\rm H}-\frac{\alpha}{2}}^{r_{\rm H} + \frac{\alpha}{2}} \rho^\prime 4\pi r^3dr .
\end{equation}
We get consecutively
\begin{equation}
	S = \frac{4\pi c^2}{3T_0}\rho_0 \left(r_{\rm H}-\frac{\alpha}{2} \right)^3 
	+ \frac{ c^2}{T_0} \int_{r_{\rm H}-\frac{\alpha}{2}}^{r_{\rm H} + \frac{\alpha}{2}} \rho \,4\pi r^2dr .
\end{equation}
For all solutions (\ref{eq:solutions}) the second term is a constant given in equation (\ref{eq:cond_3}). We finally get
\begin{equation}\label{eq:entropy}
	S = \frac{4\pi c^2}{3}\frac{\rho_0 r_{\rm H}^3}{T_0} =  \frac{M_\bullet c^2}{T_0},
\end{equation}
where we used equation (\ref{eq:r_H}) that identifies the coincidence of cosmological and black hole event horizons.
Note that this result holds for any choice of $\alpha$ and for all solutions of the cosmological black hole spectrum (\ref{eq:solutions}). Therefore, irrespectively from the exact value of the temperature $T_0$, equation (\ref{eq:entropy}) shows that all solutions (\ref{eq:solutions}) with the same total mass-energy, correspond also to the same entropy. 

Provided $\alpha$ accounts for the quantum indeterminacy of the event horizon (\ref{eq:Delta_rH}), this Tolman temperature may be identified with the cosmological temperature $T_{\rm dS}$. In such a case, by direct substitution of the cosmological temperature in entropy (\ref{eq:entropy}),
we reach the intriguing conclusion that the fluid entropy equals the Bekenstein-Hawking entropy $S_{\rm BH}$
\begin{equation}\label{eq:S}
	S = \frac{4\pi G}{\hbar c}M_\bullet^2 = \frac{4\pi r_{\rm H}^2}{4\hbar G/c^3} \equiv S_{\rm BH},
\end{equation}
if 
\begin{equation}\label{eq:T_0}
	T_0 = T_{\rm dS} \equiv \frac{\hbar c}{2\pi r_{\rm H}} =  \frac{\hbar c^3}{4\pi G M_\bullet} \equiv 2T_{\rm BH}, 
\end{equation}
where $T_{\rm BH}$ denotes the Bekenstein-Hawking temperature.
Equation (\ref{eq:S}) suggests the interpretation of the Bekenstein-Hawking entropy as the entropy of the horizon realized as a fuzzy fluid shell. 

Assuming that the Bekenstein-Hawking entropy is a universal maximum bound, the temperature should be equal to the de Sitter temperature (to maximize the entropy) and the width $\alpha$ of the shell should not exceed the appropriate quantum fuzziness, not specified in this work. In this sense, the equation of state of matter (\ref{eq:P_r}), (\ref{eq:P_T}) for the cosmological black hole (\ref{eq:solutions}) is dictated by maximum entropy and General Relativity. It describes the smooth connection between two different vaccua.

It seems remarkable that we recover the black hole entropy for a Tolman temperature that equals the cosmological temperature, fusing effectively de Sitter and Schwarzschild horizons, considering classical relativistic fluid considerations. 
In this respect, Figures \ref{fig:spectrum}, \ref{fig:orders} describe a new type of event horizon, we call a dual horizon, that is a fusion of cosmological and black hole event horizons. 
Note that the sole quantum assumption in this calculation is that the horizon's width is fuzzy. The exact measure of quantum fuzziness $\alpha$ does not affect the result.

\section{Quasi-normal modes}

A linear perturbation analysis about the static equilibrium (\ref{eq:ds2_ansatz})-(\ref{eq:P_T}), performed in Appendix \ref{app:radial}, shows that a radial perturbation cannot develop unstable radial modes. Unless it is identical to another static equilibrium, it may however develop non-radial oscillation modes. For any compact object, the latter are categorized in two types; polar and axial \cite{chandrasekhar1998the}. In Ref. \cite{2009PhRvD..80l4047P} was argued that, similarly to the Schwarzschild black hole case, polar and axial perturbations are isospectral for an ultra-compact object with a de Sitter core and an ultra-thin shell (limit $2GM \rightarrow Rc^2$ as in our case $\varepsilon \ll 1$). That is because the master equation for polar perturbations is continuous across the shell. Here we shall calculate the quasi-normal modes of axial perturbations.

\begin{figure}[!tb]
	\begin{center}
		\begin{subfigure}{0.45\textwidth}
			\includegraphics[scale = 0.5]{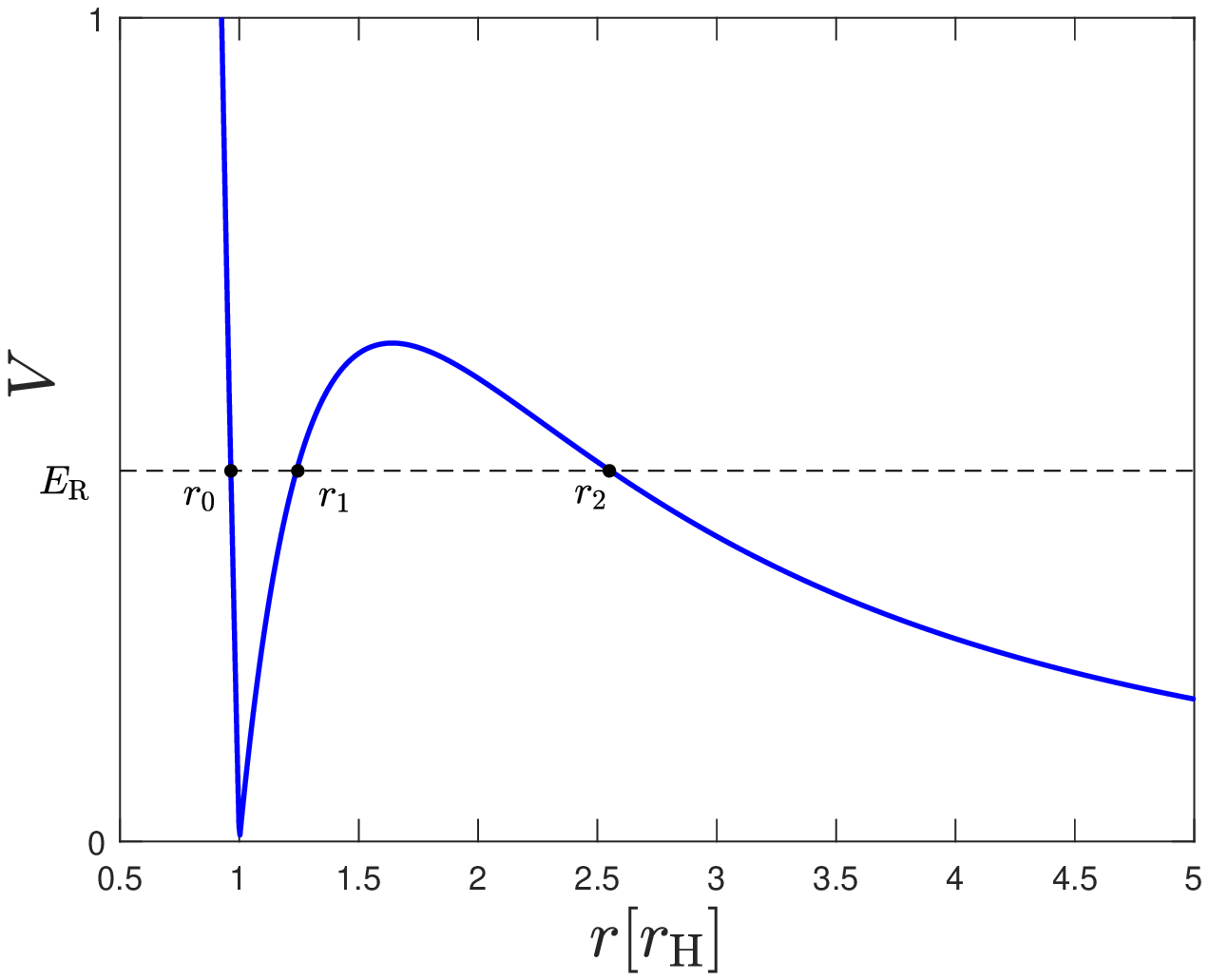}  
			%						\caption{}
			\label{fig:V_r}
		\end{subfigure}
		\begin{subfigure}{0.45\textwidth}
			\includegraphics[scale = 0.5]{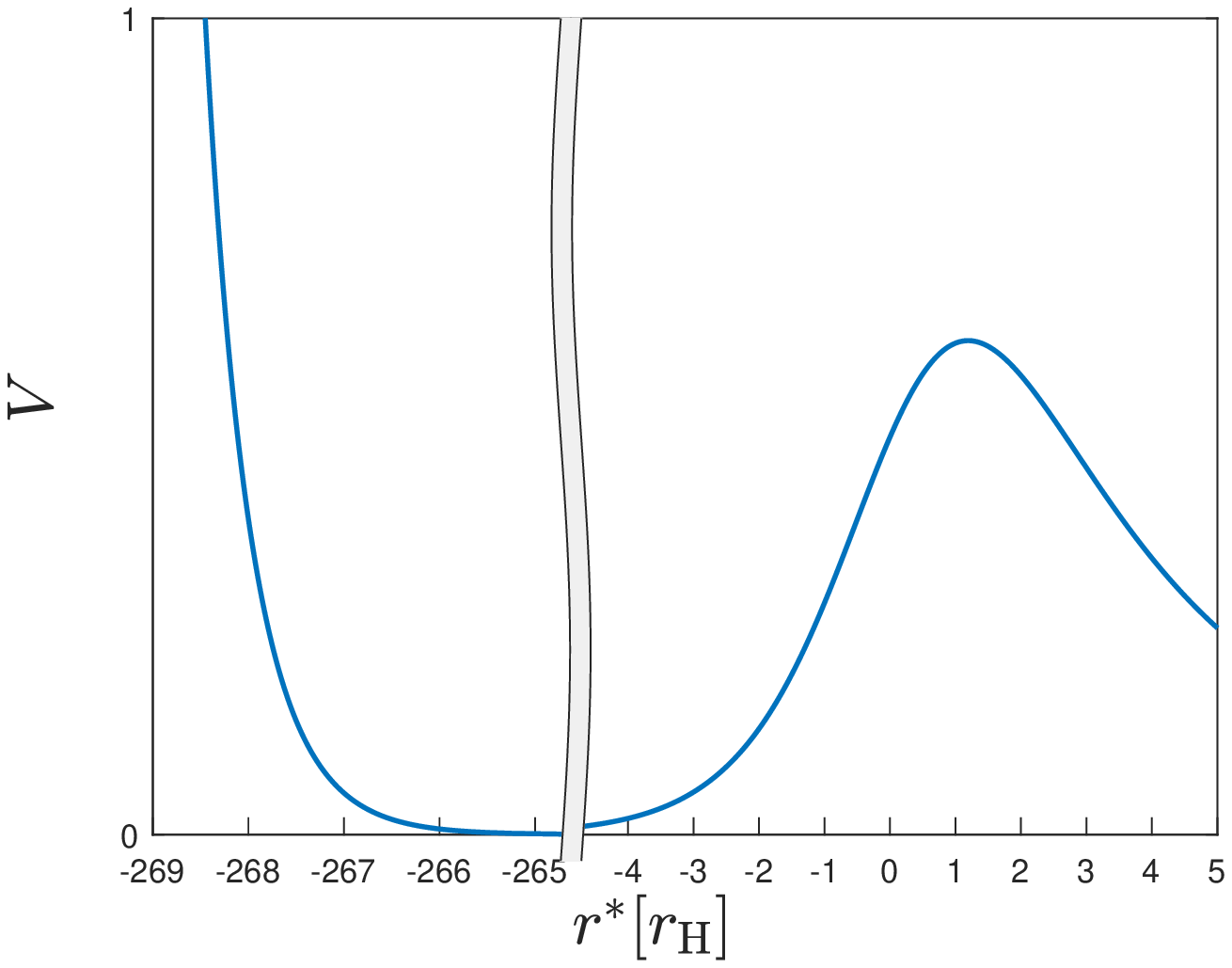}  
			%						\caption{Tortoise coordinate.}
			\label{fig:V_tortoise}
		\end{subfigure}
	\captionsetup{justification=raggedright}
		\caption{The scattering potential (\ref{eq:V}) of axial perturbations for $\alpha$ equal to the Compton wavelength of  a cosmological black hole with $M_\bullet = 10{\rm M}_\odot$, that corresponds to $\varepsilon = 3.8\cdot 10^{-78}$. We used in these plots the solution $K=1$, $N=3$ with $A_3^{(\pm)} = 8$. The radii $r_0$, $r_1$, $r_2$ are specified by the condition $E_{\rm R} = V$ and $E_{\rm R}$ is defined for each mode $n$ such that the generalized Bohr-Sommerfeld rule (\ref{eq:Bohr_Sommerfeld_R}) is satisfied. On the right panel we plot the potential with respect to the tortoise coordinate $r^*$ (\ref{eq:r_tortoise}).
			It is $r^*(r=r_{\rm H} - \alpha/2) = -180.3 r_{\rm H}$, $r^*(r=r_{\rm H}) = -179.4 r_{\rm H}$, $r^*(r=r_{\rm H} + \alpha/2) = -177.9 r_{\rm H}$.}
		\label{fig:V}
	\end{center}
\end{figure}

Axial perturbations $\psi_\ell(r,t) = e^{-i \omega  t}\phi_\ell (r)$ about the static spacetime (\ref{eq:ds2_ansatz}), (\ref{eq:dmdr}) for any metric function $h(r)$ are described by the Regee-Wheeler type of equation \cite{Dymnikova_arXiv_2005}
\begin{equation}\label{eq:SL}
	\frac{d^2\phi_\ell}{dr^{*2}} + (\frac{\omega^2}{c^2}  - V)\phi_\ell = 0,
\end{equation}
where the scattering potential is
\begin{equation}\label{eq:V}
	V(r) = h(r)\left(\frac{\ell(\ell+1)}{r^2} + \frac{8\pi G}{c^2} \rho (r) - \frac{6Gm(r)}{c^2 r^3} \right),\quad \forall r\geq 0,
\end{equation}
and $r^*$ is the, so-called, tortoise coordinate defined as
\begin{equation}\label{eq:r_tortoise}
	d r^* = \frac{1}{h(r)} dr, \quad \forall r \geq 0 .
\end{equation}
We determine the constants of integration  (see Appendix \ref{app:tortoise}) by setting $r^* = r + r_{\rm H}\ln (r/r_{\rm H}-1)$ for $r\geq r_{\rm H} + \alpha/2$ and requiring that $r^*$ is continuous at $r = r_{\rm H} \pm \alpha/2$.
The scattering potential, plotted in Figure \ref{fig:V}, is strictly positive certifying the stability of the solutions (\ref{eq:solutions}) against axial perturbations.

\begin{figure}[tb!]
	\begin{center}
		\begin{subfigure}{0.45\textwidth}
			\includegraphics[scale = 0.5]{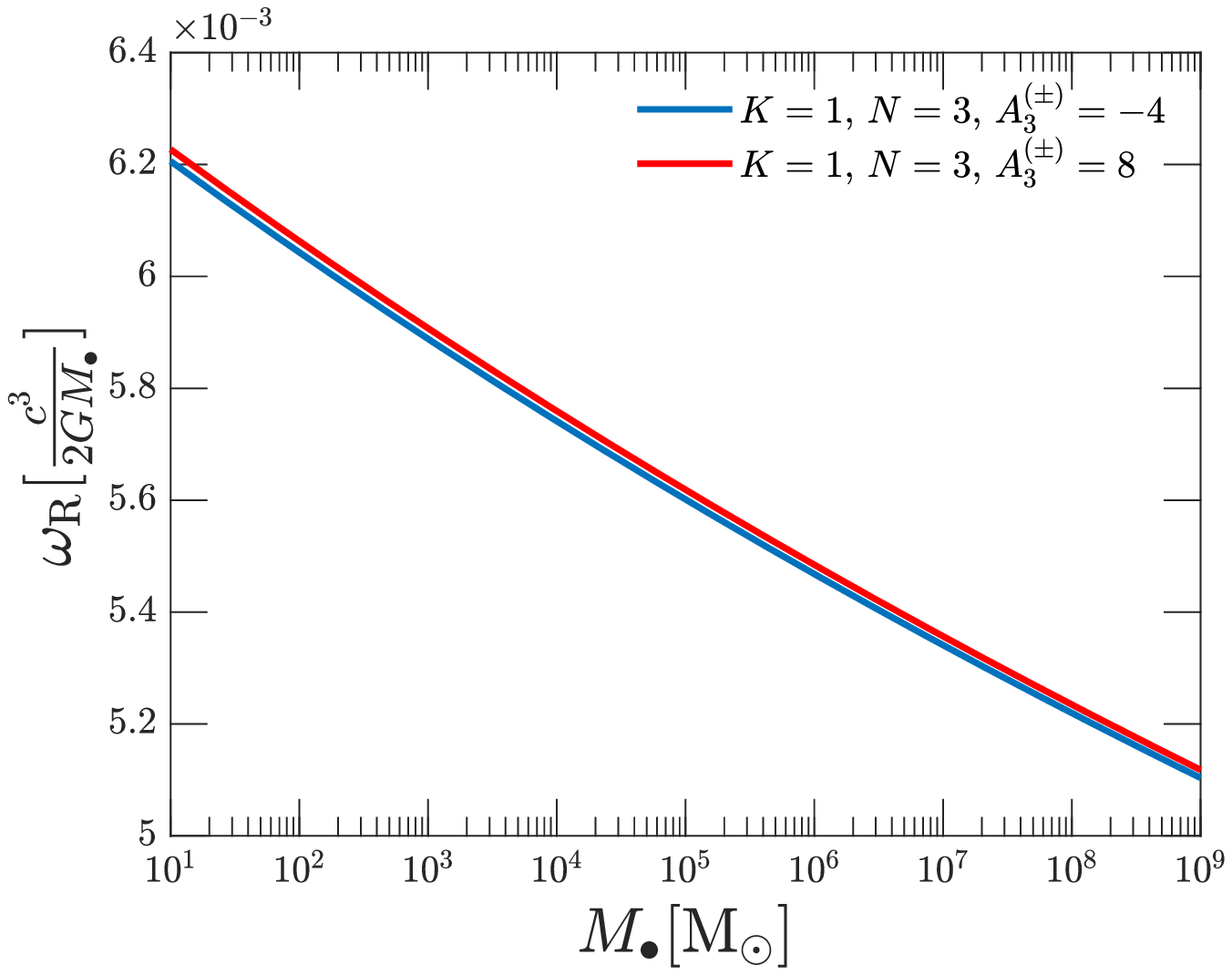}  
			%			\caption{$\omega_{\rm R}$.}
			\label{fig:omegaR_M_dimensionless_n=0_l=2}
		\end{subfigure}
		\begin{subfigure}{0.45\textwidth}
			\includegraphics[scale = 0.5]{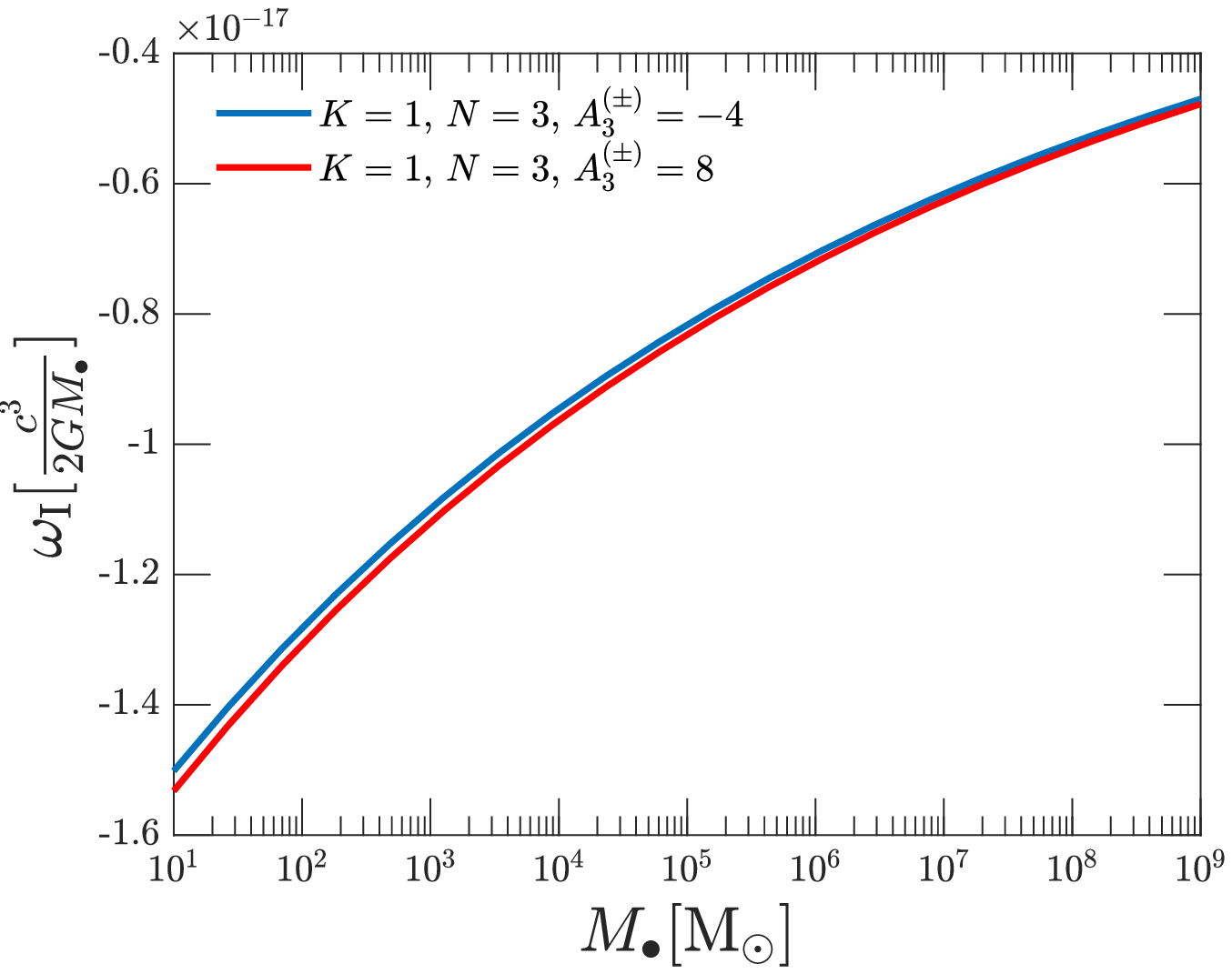}  
			%			\caption{$\omega_{\rm I}$.}
			\label{fig:omegaI_M_dimensionless_n=0_l=2}
		\end{subfigure}
	\captionsetup{justification=raggedright}
		\caption{ The real (left panel) and imaginary (right panel) frequency of the quasinormal mode $n=0$, $\ell=2$ of the cosmological black hole with respect to its mass $M_\bullet$, in dimensions $c^3/2GM_\bullet$. The two lines correspond to the two limiting solutions $A_3^{(\pm)} = 8$ (red) and $A_3^{(\pm)} = -4$ (blue) of the case $\{K=1,N=3\}$, depicted in Figure \ref{fig:rho_1st_order_spectrum}. The quasinormal modes' values of these two solutions bound all values of solutions $\{K=1,N=3\}$, $\{K=2,N=5\}$. There is also strong numerical evidence that they bound all solutions (\ref{eq:solutions}). Differences in the precise value of $\varepsilon$ are negligible.}
		\label{fig:omega_n=0_l=2}
	\end{center}
\end{figure}

\begin{table}[h]
	\begin{center}
		\begin{subtable}[]{0.45\textwidth}
			\begin{tabular}{c | c c c }
				\diagbox[width=2em,height=2em,trim=l]{$n$}{$\ell$}	& 2 & 3 & 4 
				\\
				\midrule
				0 &  0.0062 &  0.0063 &  0.0063  \\
				1 &  0.0184 &  0.0185 &  0.0186  \\
				2 &  0.0306 &  0.0307 &  0.0308  \\
				3 &  0.0426 &  0.0428 &  0.0430  \\
				4 &  0.0546 &  0.0549 &  0.0551  \\
				5 &  0.0666 &  0.0670 &  0.0672  \\
				6 &  0.0786 &  0.0790 &  0.0793  \\
				7 &  0.0905 &  0.0910 &  0.0913  
			\end{tabular}
			\caption{$\frac{2GM_\bullet}{c^3} \omega_{\rm R} $}
		\end{subtable}
		\begin{subtable}[]{0.45\textwidth}
			\begin{tabular}{c | c c c}
				\diagbox[width=2em,height=2em,trim=l]{$n$}{$\ell$}	&
				2 & 3 & 4 
				\\
				\midrule
				0   &       -1.5323{\rm E}-17 & -3.5931{\rm E}-25 & -8.2167{\rm E}-33 \\
				1	&		-3.4718{\rm E}-15 & -7.3691{\rm E}-22 & -1.5037{\rm E}-28 \\
				2	&		-4.6016{\rm E}-14 & -2.7202{\rm E}-20 & -1.5358{\rm E}-26 \\
				3	&		-2.6282{\rm E}-13 & -3.0498{\rm E}-19 & -3.3658{\rm E}-25 \\
				4	&		-9.9518{\rm E}-13 & -1.9108{\rm E}-18 & -3.4785{\rm E}-24 \\
				5	&		-2.9517{\rm E}-12 & -8.4702{\rm E}-18 & -2.2957{\rm E}-23 \\
				6	&		-7.4453{\rm E}-12 & -2.9846{\rm E}-17 & -1.1298{\rm E}-22 \\
				7	&		-1.6734{\rm E}-11 & -8.9303{\rm E}-17 & -4.4937{\rm E}-22 
			\end{tabular}
			\caption{$\frac{2GM_\bullet}{c^3} \omega_{\rm I} $}
		\end{subtable}
	\captionsetup{justification=raggedright}
		\caption{The real (left panel) and imaginary (right panel) frequency of the quasinormal modes of a cosmological black hole $M_\bullet = 10M_\odot$. These frequencies apply to the solution $A_3^{(\pm)} = 8$ of the case $\{K=1,N=3\}$. Differences between different solutions are minor, as depicted in Figure \ref{fig:omega_n=0_l=2} for $n=0$.}
		\label{tab:omega}
	\end{center}
\end{table}

\begin{figure}[!tb]
	\begin{center}
		\begin{subfigure}{0.45\textwidth}
			\includegraphics[scale = 0.5]{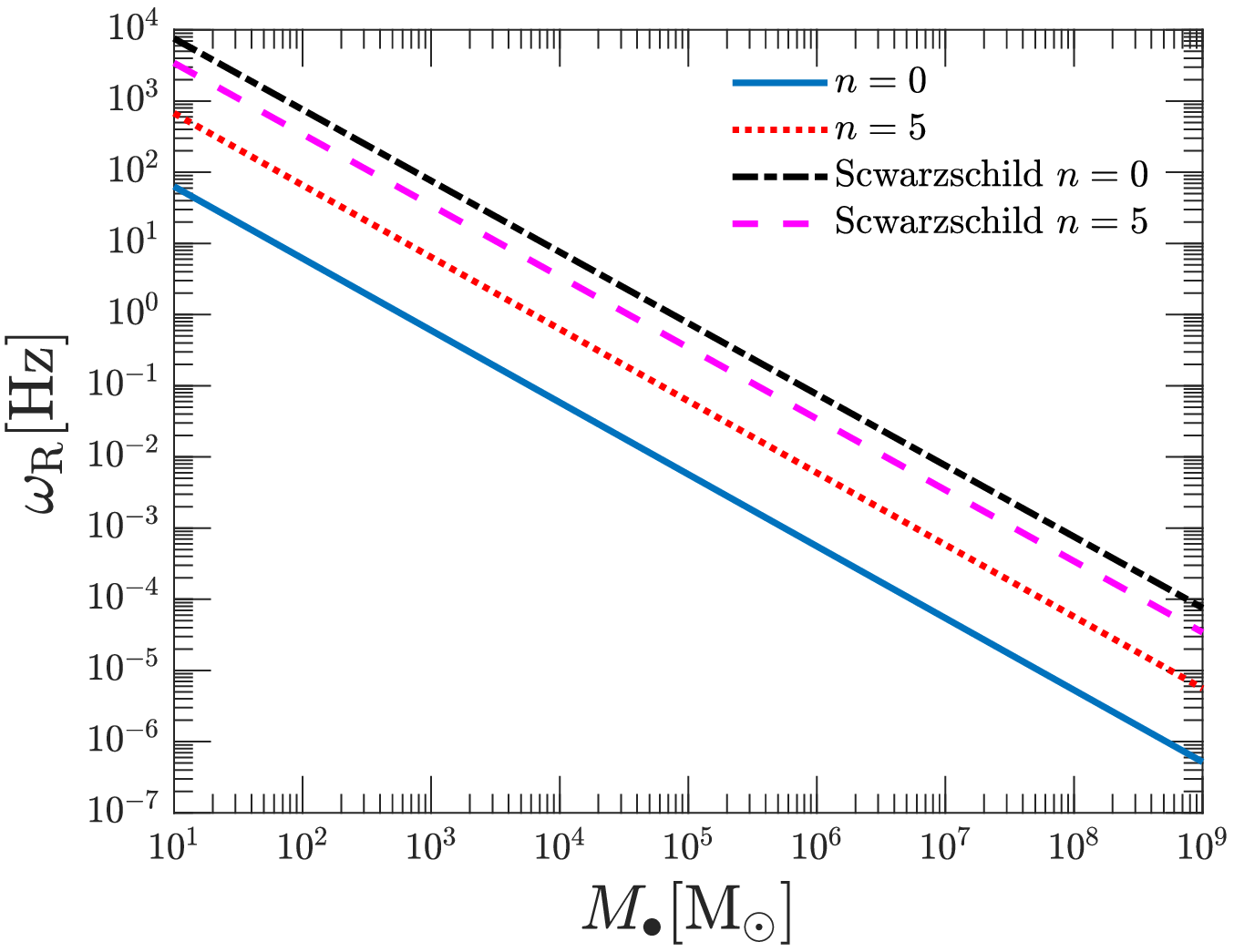}  
			%			\caption{$\omega_{\rm R}$.}
			\label{fig:omegaR_M_Hz_l=2}
		\end{subfigure}
		\begin{subfigure}{0.45\textwidth}
			\includegraphics[scale = 0.5]{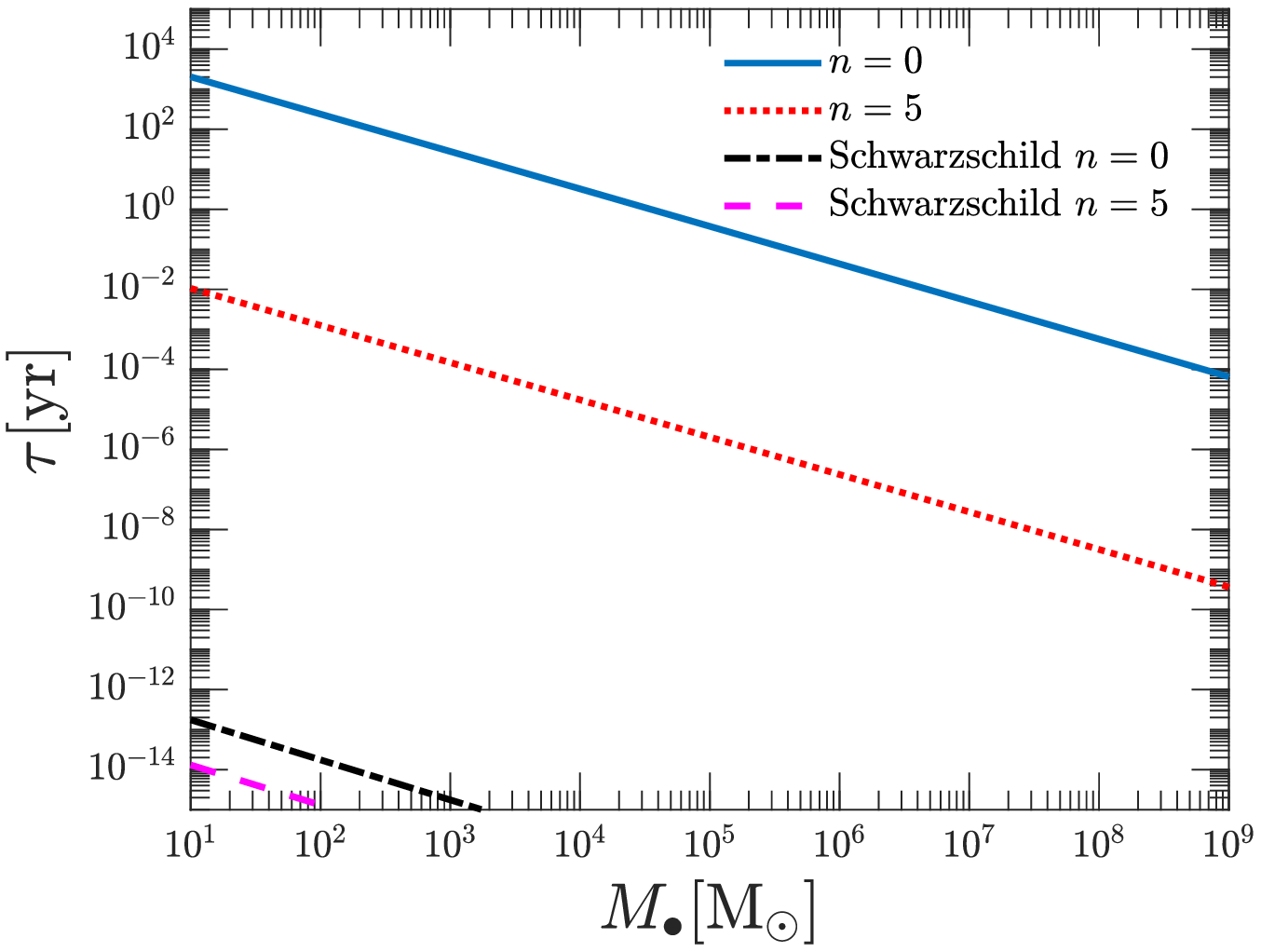}  
			%			\caption{$\tau$.}
			\label{fig:tau_M_yr_l=2}
		\end{subfigure}
	\captionsetup{justification=raggedright}
		\caption{The frequency (left panel) and damping period (right panel) of the quasinormal modes $n=0$, $n=5$  for $\ell=2$ of the cosmological black hole with respect to its mass $M_\bullet$. These apply to all solutions $\{K=1,N=3\}$, $\{K=2,N=5\}$ depicted in Figure \ref{fig:spectrum}, since any differences are negligible in the scales used. Any larger value of $K$ we tried ($\leq 15$) gives also identical results. Differences in the precise value of $\varepsilon$ are also negligible. There are also depicted the values of Schwarzschild black holes for comparison.}
		\label{fig:omega-tau_l=2}
	\end{center}
\end{figure}

The Sturm-Liouville problem (\ref{eq:SL}) can by solved (calculation of the quasi-normal modes $\omega_n$) by use of the generalized Bohr-Sommerfeld method \cite{1991PhLA..157..185P}. 
It has been shown to be sufficiently accurate in the case of ultra-compact gravastars \cite{2017CQGra..34l5006V}.
This method dictates identifying $E_{{\rm R},n}$ for some $n$ such that (see Appendix \ref{app:bohr-sommerfeld})
\begin{equation}
	\label{eq:Bohr_Sommerfeld_R}
	\int_{r^{*}_0(E_{{\rm R}, n})}^{r^*_1(E_{{\rm R}, n})} \sqrt{E_{{\rm R}, n} - V(r^*)} dr^* = \pi \left( n + \frac	{1}{2}\right).
\end{equation}
The quasi-normal mode frequencies are then specified directly as 
\begin{equation}
	\label{eq:omega}
	\omega_n = \omega_{{\rm R},n} + i \omega_{{\rm I},n} = c\sqrt{E_{{\rm R},n} + i E_{{\rm I},n}}	,
\end{equation}
where
\begin{equation} 
		\label{eq:Bohr_Sommerfeld_I}
	E_{{\rm I}, n} = - \frac{1}{2} \exp\left(-2\int_{r^*_1(E_{{\rm R}, n})}^{r^*_2(E_{{\rm R}, n})} \sqrt{V(r^*) - E_{{\rm R}, n}} dr^*\right)
	\left(\int_{r^{*}_0(E_{{\rm R}, n})}^{r^*_1(E_{{\rm R}, n})} \frac{1}{\sqrt{E_{{\rm R}, n} - V(r^*)}} dr^* \right)^{-1} .
\end{equation}
The quantities $r_0^*$, $r_1^*$ corresponding to some $r_0$, $r_1$, respectively, are the roots of the equation $E_{\rm R} - V(r) =0$ that define the bounding region $E_{\rm R} \geq V(r)$. The $r_2$ along with $r_1$ define the reflecting region $E_{\rm R} \leq V(r)$. This is depicted in Figure \ref{fig:V}. We were able to calculate the modes $\omega_{n,\ell}$ by use of mixed analytical and numerical calculations, which we describe in detail in Appendix \ref{app:bohr-sommerfeld}. 

In Figure \ref{fig:omega_n=0_l=2} we depict, for black hole masses $M_\bullet \in {[10,10^9]} M_\odot$, the frequencies of the mode $n=0$, $\ell = 2$ for the two limiting solutions of Figures \ref{fig:rho_1st_order_spectrum}, \ref{fig:P_T_1st_order_spectrum}. These correspond to $A_3^{\pm} = -4$ and $A_3^{\pm} = 8$ for $\{K=1,N=3 \}$. 
The differences between the two solutions in the values of $\omega_{\rm R}$, $\omega_{\rm I}$ are very small and decrease with increasing black hole mass. We find numerically that the two solutions bound the values of the modes for all solutions with $K=1$, $K=2$ and we find strong numerical evidence that this is true at least up to $K=15$. Our results are identical for $\alpha$ ranging from $10^9$ times the Planck length down to $\alpha$ equal to the Compton wavelength. Note that in contrast to the case of a Shwarzschild black hole, the value of $(2GM_\bullet/c^3)\omega$ depends on the cosmological black hole mass, although this dependence is very small. 

In Table \ref{tab:omega} we list the values of the quasi-normal mode frequencies for $0\leq n\leq 7$ (the importance of the first seven quasi-normal modes has been emphasized recently \cite{2019PhRvX...9d1060G,2020PhRvD.101d4033B,2021PhRvD.103j4048D,2021PhRvD.103h4048F,2021arXiv210909757O}) and $2\leq \ell \leq 4$ for a cosmological black hole with $M_\bullet = 10M_\odot$. 
For black hole masses $M_\bullet \in {[10,10^9]}{\rm M}_\odot$, the fundamental frequency lies in the range $10^{-6}{\rm Hz}\lesssim \omega_{{\rm R},0} \lesssim 50{\rm Hz}$ and the $n=5$ overtone is about ten times bigger, as depicted in Figure \ref{fig:omega-tau_l=2}. The damping time of the overtones are on the other hand drastically lower with respect to the fundamental mode. 

The fundametal mode alone may be used to reconstruct the full inspiral merger ringdown waveform of a binary black hole merger signal (e.g. GW150914 \cite{PhysRevLett.116.221101,PhysRevLett.121.129902}). The highest possible fundamental mode of an astrophysical  cosmological black hole (corresponding to the minimum possible mass $M_\bullet = 10{\rm M}_\odot$, that is a merger of two $5{\rm M}_\bullet$ black holes) is $63{\rm Hz}$ (Figure \ref{fig:omega-tau_l=2}) which lies outside the detection range of LIGO-Virgo. This does not mean that LIGO-Virgo observations exclude the possibility that they involved cosmological black holes, but only that LIGO-Virgo cannot discrimininate between a singular (Schwarzschild or Kerr) and a cosmological black hole. The reason is the, so called, ``mode camouflage'' mechanism \cite{2014PhRvD..90d4069C}. The ringdown modes of a black hole (regular or not) are determined by the external null geodesic and not by interior fluctuations \cite{PhysRevD.79.064016}. Fluctuations generated inside our cosmological black hole will dominate after the exterior perturbations are damped. Thus, LIGO-Virgo cannot discriminate between a Schwarzschild (or Kerr) black hole and a cosmological black hole. Following the full damping of the external light-ring modes, the internal fluctuation frequencies lie outside the frequency range detectability of LIGO-Virgo. 

However, it is evident from Figure \ref{fig:omega-tau_l=2}  that for $M_\bullet \gtrsim 10^4{\rm M}_\odot$ the fundamental mode of cosmological black hole fluctuations lies within the frequency detectability range ($\sim 10^{-1}-10^{-5}{\rm Hz}$) of the LISA space interferometer. It sounds in particular intriguing that LISA may be able to detect an intermediate mass cosmological black hole through its postmerger ringdown phase, even if the binary inspiral phase cannot be detected.
Still, in order to estimate the minimum possible amplitude sensitivity of an interferometer so as to detect a cosmological black hole ringdown, the excitation factors of its quasi-normal modes, following a binary merger, have to be calculated. This is an involved task, that this work urges the community to perform.
In every case, our results as in Figure \ref{fig:omega-tau_l=2}, clearly suggest that despite the well-known mode camouflage mechanism of ultra-compact objects \cite{2014PhRvD..90d4069C} mentioned above, cosmological black holes particularly are in principle detectable and distinguishable from singular black holes. If already LISA is not amplitude-wise sensitive enough, it is a matter of developing the appropriate technology to detect cosmological black holes provided they exist.

Finally, let us remark that regular black holes, like the one we propopse, should not suffer from the instabilities, such as the light-ring instability \cite{Cunha_2017,2014PhRvD..90d4069C}, the ergosphere instability \cite{2008PhRvD..77l4044C}
and the accretion instability \cite{2018PhRvD..97l3012C,2019arXiv190208180C,2020EPJC...80...36A}, that have been argued to occur in gravastars and dark energy stars. The absence of a horizon is a key assumption that drives the appearance of these instabilities \cite{2019PhRvD..99f4007M}. 

\section{Discussion}

We discover that in General Relativity, regular black holes containing a de Sitter core correspond to a spectrum of spacetime solutions assuming quantum indeterminacy of the localization of the horizon, which behaves as an anisotropic fluid shell. 
All spacetime states of the cosmological black hole spectrum have the same energy and entropy, resembling a quantum degeneracy. This is a fluid entropy. It recovers the Bekenstein-Hawking black hole entropy if the Tolman temperature of the fluid is identified with the temperature of the cosmological horizon, fusing the cosmological and black hole horizons in a single dual horizon. 

The quasi-normal modes of cosmological black holes are distinctively different than the ones of Schwarzschild black holes. Still, LIGO-Virgo cannot disciminate between cosmological and Schwarzschild black holes, because of the well-known mode camouflage mechanism --ringdown waveform is dominated initially by spacetime fluctuations in the region of the external null geodesic, that is common in regular and singular black holes-- and the fact that the mode frequencies of astrophysical cosmological black holes, namely $10^{-6}{\rm Hz} \lesssim \omega_{\rm R} \lesssim 10{\rm Hz}$, lie outside the frequency's range detectability of LIGO-Virgo. Therefore, it remains open the possibility that LIGO-Virgo's detections are cosmological black holes. Most importantly, the quasi-normal frequency range of astrophysical cosmological black holes lies inside the detectability frequency range of the planned space interferometer LISA. Thus, this work urges the community to investigate further the properties of cosmological black holes, proposed here, and especially their inspiral and ringdown waveforms. There arises the fascinating possibility that black hole detections are also detections of dark energy universes. If these may evolve to inflationary universes similar to our own and if the latter is itself such an object remain open possibilities that beg for further investigation.

\bibliography{DEBH_2021}

\appendix

\section{Derivation of the Poisson-Israel solution}\label{app:P-I} 

Assuming the static, spherically symmetric ansatz (\ref{eq:ds2_ansatz}), the Einstein equations 
\begin{equation}\label{eq:einstein}
	R_\mu^{~\nu} - \frac{1}{2} R_\sigma^{~\sigma} \delta_\mu^{~\nu} = \frac{8\pi G}{c^4}T_\mu^{~\nu}
\end{equation}
give
\begin{align}
	\label{eq:einstein_r}
	\frac{8\pi G}{c^4} T_0^{~0} = \frac{8\pi G}{c^4} T_1^{~1} &=
	\frac{1}{r} h^\prime + \frac{1}{r^2} h - \frac{1}{r^2} ,
	\\
	\label{eq:einstein_T}
	\frac{8\pi G}{c^4} T_2^{~2} = \frac{8\pi G}{c^4} T_3^{~3} &=
	\frac{1}{2} h^{\prime\prime} + \frac{1}{r} h^\prime, 
\end{align}
where prime denotes differentiation with respect to $r$. Let us denote $T_0^{~0} = -\rho(r) c^2$, $T_1^{~1} = P_{\rm r} (r)$, $T_2^{~2}=T_3^{~3} = P_{\rm T}(r)$, where $\rho(r)$ has dimensions of mass density and $P_{\rm r}(r)$, $P_{\rm T}(r)$ dimensions of pressure. 

Assuming further a function $m(r)$ such that
\begin{equation}\label{eq:h}
	h(r) = 1 - \frac{2 G m(r)}{r c^2},
\end{equation}
the Einstein equations (\ref{eq:einstein_r}), (\ref{eq:einstein_T}) give
\begin{align}
	\label{eq:dm_methods}
	\frac{dm}{dr} &= 4\pi r^2 \rho(r), \\
	\label{eq:P_r_methods}
	P_{\rm r}(r) &= - \rho(r) c^2, \\
	\label{eq:P_T_methods}
	P_{\rm T}(r) &= - \rho(r) c^2 - \frac{1}{2} r \rho^\prime (r)	
\end{align}

One solution of these equations is
\begin{align}
	\label{eq:m_PI}
	m^{(\rm PI)}(r) &= M_\bullet \left\lbrace \theta \left(\frac{r}{r_{\rm H}} - 1\right) + \frac{r^3}{r_{\rm H}^3} \left(\theta \left(\frac{r}{r_{\rm H}}\right) - \theta \left(\frac{r}{r_{\rm H}} - 1\right) \right)\right\rbrace 
	\\
	\label{eq:rho_PI}
	\rho^{(\rm PI)}(r) &= \rho_0 \left\lbrace\theta \left(\frac{r}{r_{\rm H}}\right) - \theta \left(\frac{r}{r_{\rm H}} - 1\right) \right\rbrace , \\
	\label{eq:P_r_PI}
	P^{(\rm PI)}_{\rm r}(r) &= - \rho^{(\rm PI)}(r) c^2 ,\\
	\label{eq:P_T_PI}
	P^{(\rm PI)}_{\rm T}(r) &= -\rho^{(\rm PI)}(r) c^2 + \frac{1}{2} \rho_0 \delta \left(\frac{r}{r_{\rm H}} - 1 \right),
\end{align}
where $\rho_0 = M_\bullet/(4\pi r_{\rm H}^3/3)$, $r_{\rm H} = 2GM_\bullet/c^2$ and $M_\bullet = m^{(\rm PI)}(r_{\rm H})$. 
We denote $\delta (x)$ the Dirac $\delta$-function and $\theta(x)$ the Heaviside step function
\begin{equation}
	\theta (x)	 = \left\lbrace 
	\begin{array}{ll}
		1 &,\, x\geq 0\\
		0 &,\, x < 0	
	\end{array}
	\right.
\end{equation}

The superscript (PI) is an acronym for ``Poisson-Israel'', because the expression (\ref{eq:P_T_methods}) has appeared for the first time, to our knowledge, in Ref. \cite{1988CQGra...5L.201P}. Poisson \& Israel remarked that an observer at a proper distance $\Delta s$ outside the horizon will perceive an infinite tangential pressure on the horizon
\begin{equation}\label{eq:P_T_divergence}
	P^{({\rm PI})}_{\rm T} = \frac{1}{2}\rho_0 \delta\left(g_{rr}^{-1/2} \frac{\Delta s}{r_{\rm H}} \right) = \frac{1}{2}\rho_0 r_{\rm H}  g_{rr}^{1/2} \delta(\Delta s)
	\rightarrow \infty, \;\text{for} \; r = r_{\rm H}.
\end{equation}

Let us now prove that equations (\ref{eq:m_PI})-(\ref{eq:P_T_PI}) satisfy the Einstein equations (\ref{eq:dm_methods})-(\ref{eq:P_T_methods}).
We shall use the dimensionless variable
\begin{equation}
	u = \frac{r}{r_H},
\end{equation}
and the dimensionless quantities
\begin{equation}\label{eq:nond}
	\tilde{m} = \frac{m}{M_\bullet},\;
	\tilde{\rho} = \frac{\rho}{\rho_0},\;
	\tilde{P}_{\rm r} = \frac{P_{\rm r}}{\rho_0 c^2},\;
	\tilde{P}_{\rm T} = \frac{P_{\rm T}}{\rho_0 c^2},	
\end{equation}
The dimensionless function $\tilde{m}^{(\rm PI)}(u)$ is written as
\begin{equation}
	\tilde{m}^{(\rm PI)}(u) = \theta \left(u - 1\right) + u^3 \left(\theta \left(u\right) - \theta \left(u - 1\right) \right).
\end{equation}
Using the property $\theta^\prime (u)= \delta(u)$ we have that
\begin{align}
	\frac{d \tilde{m}^{(\rm PI)}}{d u} &= \delta (u-1) + u^3(\delta (u) - \delta (u-1)) + 3u^2 (\theta(u) - \theta (u-1)) \\
	&= u^2 u\delta(u) -(u^2 + u + 1) (u-1) \delta (u-1) + 3u^2 (\theta(u) - \theta (u-1)),
\end{align}
which, considering the identity $u\delta(u) = 0$, gives
\begin{equation}\label{eq:dmdu}
	\frac{d \tilde{m}^{(\rm PI)}}{d u} = 3 u^2 (\theta(u) - \theta (u-1)).
\end{equation}
Comparing with (\ref{eq:dm_methods}) we get
\begin{equation}
	\tilde{\rho}^{(\rm PI)} = \theta(u) - \theta (u-1), 
\end{equation}
that is equation (\ref{eq:rho_PI}).

Now, we have that
\begin{equation}
	u(\tilde{\rho}^{(\rm PI)}(u))^\prime  = u\delta(u) - u\delta(u-1) = u\delta(u-1) = (x+1)\delta(x) = \delta(x) = \delta(u-1)
\end{equation}
where we have used the transformation $x = u-1$. Thus, the tangential pressure (\ref{eq:P_T_methods}) is
\begin{equation}
	\tilde{P}^{(\rm PI)}_{\rm T}(u) = -\tilde{\rho}^{(\rm PI)}(u) + \frac{1}{2}\delta(u-1),
\end{equation}
that is the Poisson-Israel solution (\ref{eq:P_T_PI}).

\section{Analytical expressions of the cosmological black hole spectrum}\label{app:DE-BH}

We shall use in the followings the dimensionless quantities (\ref{eq:x_nond}), (\ref{eq:nond}).
The spectrum (\ref{eq:solutions}) is written in these dimensionless variables as
\begin{equation} 
	\label{eq:solutions_nond}
	\tilde{\rho}(x) = \left\lbrace
	\begin{array}{ll}
		1 &, \; x \leq - \frac{1}{2}, \\
		\tilde{\rho}_{(-)}(x) &, \;   - \frac{1}{2} \leq x \leq 0, \\
		\tilde{\rho}_{(+)}(x) &, \;  0 \leq x \leq \frac{1}{2} , \\
		0 &, \; x \geq \frac{1}{2}. \\
	\end{array}
	\right. 
	,
	\quad	
	\tilde{\rho}_{(\pm)}(x) = \sum_{n=0}^N A^{(\pm)}_n (\varepsilon) x^n .
\end{equation} 
For $K=1$ and $N=3$ the conditions (\ref{eq:cond_1})-(\ref{eq:cond_3}) give 
\begin{align}
	A_0^{(-)} &= A_0^{(+)} = \frac{ 60 - (18  + B )\varepsilon + 2 \varepsilon^2}{4 (\varepsilon^2 - 4 \varepsilon + 30)} = \frac{1}{2} + \mathscr{O}(\varepsilon),
	\\
	A_1^{(-)} &= A_1^{(+)} = \frac{-240 + 30 B + 72\varepsilon + (B - 8)\varepsilon^2 }{4(\varepsilon^2 - 4\varepsilon + 30)} = \frac{B-8}{4} + \mathscr{O}(\varepsilon),
	\\
	A_2^{(-)} &= \frac{-60 + 30B + (78+ 3B)\varepsilon + (B - 2)\varepsilon^2 }{\varepsilon^2 - 4\varepsilon + 30} = B-2 + \mathscr{O}(\varepsilon),
	\\
	A_2^{(+)} &= -\frac{-60 + 30B + (18 - 3B)\varepsilon + (B - 2)\varepsilon^2}{\varepsilon^2 - 4\varepsilon + 30} = -(B-2) + \mathscr{O}(\varepsilon),
	\\
	A_3^{(-)} &=  \frac{30B + (80 + 4B)\varepsilon + B\varepsilon^2}{\varepsilon^2 - 4\varepsilon + 30} = B + \mathscr{O}(\varepsilon),
	\\
	A_3^{(+)} &=  B 	
\end{align}
The additional requirement 
\begin{equation}
	\rho^\prime \leq 0 	
\end{equation}
imposes the constraint, to zero-th order of $\varepsilon$,
\begin{equation}
	-4 \leq B \leq 8 .
\end{equation} 

For $K=2$ and $N=5$ we get 
\begin{align}
	A_0^{(-)} &= A_0^{(+)} = \frac{129024 +168( B_2 - B_1) + ( - 21504 + 128(B_1 + B_2))\varepsilon + (3072+9(B_2-B_1))\varepsilon^2 }{6144(\varepsilon^2 + 42)} 
	\nonumber \\
	&= \frac{1}{2} + \frac{B_2 - B_1}{1536} + \mathscr{O}(\varepsilon),
	\\
	A_1^{(-)} &= A_1^{(+)} = - \left( \frac{B_1+B_2}{96} + 2\right),
	\\
	A_2^{(-)} &= A_2^{(+)} = -\frac{840 (B_2 - B_1) + (- 21504 + 128 (B_1 + B_2)) \varepsilon + 25 (B_2 - B_1)\varepsilon^2}{256 (\varepsilon^2 + 42)} 
	\nonumber \\
	&= -\frac{5}{64}(B_2-B_1) + \mathscr{O}(\varepsilon),
	\\
	A_3^{(-)} &= \frac{672(48 + 4B_1 - B_2) + 128(168 - B_1 - B_2)\varepsilon + (69B_1 - 21 B_2 + 768)\varepsilon^2 }{96(\varepsilon^2 + 42)} \nonumber \\
	&= \frac{1}{6}(48 + 4B_1 - B_2) + \mathscr{O}(\varepsilon)
	,
	\\
	A_3^{(+)} &= \frac{672(48 - B_1 - 4B_2) + 128(168 + B_1  + B_2 )\varepsilon + (- 21 B_1 + 69 B_2 + 768) \varepsilon^2}{96 (\varepsilon^2 + 42)}, \nonumber \\
	&= \frac{1}{6}(48 - B_1 - 4B_2) + \mathscr{O}(\varepsilon)
	,
	\\
	A_4^{(-)} &= \frac{129024 + 23352 B_1 - 1848 B_2 + 384(168  -  B_1  - B_2) \varepsilon + (3072 + 571 B_1 - 59 B_2 ) \varepsilon^2 }{384 (\varepsilon^2 + 42)} \nonumber \\
	&= \frac{3072 + 556 B_1 - 44 B_2}{384} + \mathscr{O}(\varepsilon)
	,
	\\
	A_4^{(+)} &= -\frac{ 129024 - 1848 B_1 + 23352 B_2 - 384( 168 - B_1 - B_2) \varepsilon + (3072 - 59 B_1 + 571 B_2)\varepsilon^2}{384(\varepsilon^2 + 42)} \nonumber \\
	&= - \frac{3072 - 44 B_1 + 556 B_2 }{384} + \mathscr{O}(\varepsilon),
	\\
	A_5^{(-)} &= B_1,
	\\
	A_5^{(+)} &= B_2
\end{align}
For $B_1=B_2=B$, the requirement $\rho^\prime \leq 0$ imposes the constraint, to zero-th order of $\varepsilon$,
\begin{equation}
	-96 \leq B \leq 24 .
\end{equation} 

Note that if we identify $\alpha$ with Compton wavelength $h/(M_\bullet c)$ then 
\begin{equation}\label{eq:Compton}
	\varepsilon_{\rm Compton} = 3.8\cdot 10^{-76}\left( \frac{M_\bullet}{{\rm M}_\odot}\right)^{-2},
\end{equation}
and if we identify it with the Planck scale, then
\begin{equation}\label{eq:Planck}
	\varepsilon_{\rm Planck} = 5.5\cdot 10^{-39} \left( \frac{M_\bullet}{{\rm M}_\odot}\right)^{-1} .
\end{equation}
The parameter $\alpha$ expresses the quantum fuzziness of the horizon. It is a free parameter within our framework and may be bigger than the Planck length.

\section{Radial perturbations}\label{app:radial}

We shall consider here radial perturbations about the static equilibrium (\ref{eq:ds2_ansatz})-(\ref{eq:P_T}).
The general energy-momentum tensor of a spherical anisotropic fluid may be written as
\begin{equation}
	T^\mu_\nu = (\rho c^2 + P_{\rm r})U^\mu U_\nu + P_{\rm r} \delta^\mu_\nu  + (P_{\rm T}- P_{\rm r})(Y^\mu Y^\nu + Z^\mu Z^\nu),
\end{equation}
where $U^\mu U_\mu = -1$, $Y^\mu Y_\mu = Z^\mu Z_\mu = 1$, $U^\mu Y_\mu = U^\mu Z_\mu = Y^\mu Z_\mu = 1$ and in co-moving coordinates $Y^\mu = \delta ^\mu_2$, $Z^\mu = \delta ^\mu_3$. The general non-static, spherically symmetric metric may be written as 
\begin{equation}
	ds^2 = - e^{\nu(r,t)} dt^2 + e^{\lambda(r,t)} dr^2 + r^2(d\theta^2 + \sin^2\theta d\phi^2 ) ,
\end{equation}
where here for convenience we suppress $c$ in $g_{00}$ identifying $t\equiv x^0$.
Assuming a spherically symmetric deviation from static equilibrium 
\begin{equation}
	u = \frac{dr}{dt} ,
\end{equation}
the non-zero components of the time-like vector $U^\mu$ are
\begin{equation}
	U^0 = \frac{e^{-\frac{\nu}{2}}}{\sqrt{1-u^2e^{\lambda - \nu}}},\;
	U^1 = \frac{u\, e^{-\frac{\nu}{2}}}{\sqrt{1-u^2e^{\lambda - \nu}}},\;
	U_0 = -\frac{e^{\frac{\nu}{2}}}{\sqrt{1-u^2e^{\lambda - \nu}}},\;
	U_1 = \frac{u\, e^{\lambda -\frac{\nu}{2}}}{\sqrt{1-u^2e^{\lambda - \nu}}}.
\end{equation}
The non-zero components of the energy momentum tensor are therefore
\begin{align}
	T^0_0 & = -\rho c^2\frac{1}{1- u ^2 e^{\lambda - \nu}} - P_{\rm r} \frac{u^2 e^{\lambda - \nu}}{1- u ^2 e^{\lambda - \nu}}
	\\
	T^1_1 &= P_{\rm r} \frac{1 }{1- u ^2 e^{\lambda - \nu}}
	+ \rho c^2 \frac{u^2 e^{\lambda - \nu}}{1- u ^2 e^{\lambda - \nu}}
	\\
	T^2_2 &= T^3_3 = P_{\rm T} 
	\\
	T^1_0 & = - (\rho c^2 + P_{\rm r}) \frac{u}{1- u ^2 e^{\lambda - \nu}}
	\\
	T^0_1 &= (\rho c^2 + P_{\rm r}) \frac{u e^{\lambda - \nu}}{1- u ^2 e^{\lambda - \nu}}.
\end{align}
The non-zero components of the Einstein tensor are
\begin{align}
	G^0_0 &= e^{-\lambda}\left( -\frac{\lambda^\prime}{r} + \frac{1}{r^2} \right) -\frac{1}{r^2}
	\\
	G^1_1 &= e^{-\lambda}\left( \frac{\nu^\prime}{r} + \frac{1}{r^2} \right) -\frac{1}{r^2}
	\\
	G^2_2 &= G^3_3 = 
	e^{-\lambda}\left( \frac{\nu^{\prime\prime}}{2} + \frac{(\nu^\prime)^2}{4} - \frac{\nu^\prime\lambda^\prime}{4} + \frac{\nu^\prime - \lambda^\prime}{2r} \right) - e^{-\nu} \left( \frac{\ddot{\lambda}}{2} + \frac{\dot{\lambda^2}}{4} - \frac{\dot{\lambda}\dot{\nu}}{4} \right)
	\\
	G^1_0 &= \frac{1}{r}e^{-\lambda} \dot{\lambda}
	\\
	G^0_1 &= - \frac{1}{r}e^{-\nu} \dot{\lambda} .
\end{align}
In our convention, the Einstein equations read
\begin{equation}
	G^\mu_\nu = \frac{8\pi G}{c^4} T^\mu_\nu .
\end{equation}
Differentiating $G^1_1$, substituting it into $G^2_2$ and using Einstein equations to substitute $G^\mu_\nu$ for $T^\mu_\nu$ we get the equation of hydrodynamic equilibrium
\begin{equation}\label{eq:cont_1}
	(T^1_1)^\prime = - \frac{\nu^\prime}{2}(T^1_1 - T^0_0) + \frac{2}{r}(T^2_2 - T^1_1) + \frac{c^4}{8\pi G r}e^{-\nu} \left( \ddot{\lambda} + \frac{\dot{\lambda^2}}{2} - \frac{\dot{\lambda}\dot{\nu}}{2} \right).
\end{equation}
We may reach to the same equation by using the continuity equation $T^\mu_{1 ; \mu} = 0$ and substituting $T^0_1$ for $G^0_1$. The time component of the continuity equation $T^\mu_{0 ; \mu}$ gives
\begin{equation}\label{eq:cont_2}
	-\dot{T^0_0} = (T^1_0)^\prime - \frac{1}{2}\dot{\lambda}(T^1_1 - T^0_0) + T^1_0 \left( \frac{\lambda^\prime+ \nu^\prime}{2} + \frac{2}{r} \right).
\end{equation}

Let us now consider the perturbations
\begin{equation}
	\lambda = \lambda_{\rm eq} + \delta \lambda,\;
	\nu = \nu_{\rm eq} + \delta \nu,\;
	\rho = \rho_{\rm eq} + \delta \rho,\;
	P_{\rm r} = P_{\rm r,eq} + \delta P_{\rm r},\;
	P_{\rm T} = P_{\rm T,eq} + \delta P_{\rm T},\;
	u = u_{\rm eq} + \delta u.
\end{equation}
Subscript ``eq'' denotes equilibrium quantities. It is
\begin{equation}
	e^{\nu_{\rm eq}} = e^{-\lambda_{\rm eq}} = h,
	\;
	P_{\rm r,eq} = -\rho_{\rm eq} c^2,\; 
	P_{\rm T,eq} = -\rho_{\rm eq} c^2 - \frac{1}{2}r \rho_{\rm eq}^\prime,\; 
	u_{\rm eq} = 0,	
\end{equation}
so that $u=\delta u$. To the first order we get
\begin{align}
	\delta^{(1)} T^0_0 &= -\delta \rho c^2,\;
	\delta^{(1)} T^1_1 = \delta P_{\rm r},\;
	\delta^{(1)} T^2_2 = \delta P_{\rm T},
	\\
	\delta^{(1)} T^1_0 &= -(\rho_{\rm eq}c^2 + P_{\rm r,eq})\delta u = 0,
	\\
	\delta^{(1)} T^0_1 &= h^{-2}(\rho_{\rm eq}c^2 + P_{\rm r,eq}) \delta u= 0,
	\\
	\delta^{(1)} G^0_0 &= -\frac{1}{r^2} \left( rh\delta\lambda^\prime + h\delta \lambda + rh^\prime \delta \lambda\right),
	\\
	\delta^{(1)} G^1_1 &= \frac{1}{r^2} \left( rh\delta\nu^\prime - h\delta \lambda - rh^\prime \delta \lambda\right) ,
	\\
	\delta^{(1)} G^2_2 &= -\delta \lambda 
	\left( \frac{h^{\prime\prime}}{2} + \frac{h^\prime}{r} \right) +
	\frac{h}{2}
	\left( \delta\nu ^{\prime\prime} + \frac{3h^\prime}{2h}\delta\nu^\prime - \frac{h^\prime}{2h}\delta\lambda^\prime + \frac{\delta\nu^\prime - \delta\lambda^\prime}{r} \right)
	+ \frac{1}{2h}\ddot{\delta\lambda} ,
	\\
	\delta^{(1)} G^1_0 &= \frac{h}{r}\dot{\delta \lambda}
	\\
	\delta^{(1)} G^0_1 &= - \frac{1}{h r} \dot{\delta \lambda} .
\end{align}
We may describe perturbations by use of $\{0,0\}$, $\{1,1\}$, $\{1,0\}$, $\{0,1\}$ components of Einstein equations along with continuity equations (\ref{eq:cont_1}), (\ref{eq:cont_2}) and get respectively
\begin{align}
	\label{eq:d1_pert_1}
	\frac{8\pi G}{c^2}\delta \rho &= \frac{1}{r^2} \left( rh\delta\lambda^\prime + h\delta \lambda + rh^\prime \delta \lambda\right),
	\\
	\label{eq:d1_pert_2}
	\frac{8\pi G}{c^4}\delta P_{\rm r} &= \frac{1}{r^2} \left( rh\delta\nu^\prime - h\delta \lambda - rh^\prime \delta \lambda\right) ,
	\\
	\label{eq:d1_pert_3}
	0 &= \frac{h}{r}\dot{\delta \lambda},
	\\
	\label{eq:d1_pert_4}
	0 &= - \frac{1}{h r} \dot{\delta \lambda} ,
	\\
	\label{eq:d1_pert_5}
	\delta P_{\rm r}^\prime &= -\frac{h^\prime}{2h} (\delta\rho c^2 + \delta P_{\rm r}) + \frac{2}{r}(\delta P_{\rm T} - \delta P_{\rm r}) + \frac{c^4}{16\pi G}\frac{1}{rh}\ddot{\delta \lambda},
	\\
	\label{eq:d1_pert_6}
	\dot{\delta \rho} c^2 &= 0 .
\end{align}
Equations (\ref{eq:d1_pert_3}), (\ref{eq:d1_pert_4}), (\ref{eq:d1_pert_6}) suggest directly that radial perturbations in the density and the radial metric component can only be static
$\dot{\delta\lambda} = \dot{\delta \rho} = 0$. We conclude that radial perturbations cannot develop unstable radial modes. If the radial perturbation is not identical to another static equilibrium state, it may develop non-radial modes.

\section{Tortoise coordinate}\label{app:tortoise}

The tortoise coordinate used in Figure \ref{fig:V} is defined as
\begin{equation}
	\frac{r^*(r)}{r_{\rm H}} = \left\lbrace
	\begin{array}{ll}
		\frac{r}{r_{\rm H}} + \ln (\frac{r}{r_{\rm H}} - 1), & \frac{r}{r_{\rm H}}\geq 1 + \frac{\varepsilon}{2} 
		\\ [2ex]
\displaystyle		R_2^* - \int_{r/r_{\rm H}}^{1 + \frac{\varepsilon}{2}} \frac{1}{h_{+}(u)} du, & 1 \leq \frac{r}{r_{\rm H}} \leq 1 + \frac{\varepsilon}{2} 
		\\ [2ex]
\displaystyle		R_H^* - \int_{r/r_{\rm H}}^{1 - \frac{\varepsilon}{2}} \frac{1}{h_{-}(u)} du, & 1 - \frac{\varepsilon}{2} \leq \frac{r}{r_{\rm H}} \leq 1 
		\\ [2ex]
		R_1^* - {\rm atanh}(1-\frac{\varepsilon}{2}) + {\rm atanh}(\frac{r}{r_{\rm H}}), & 0 \leq \frac{r}{r_{\rm H}} \leq 1 - \frac{\varepsilon}{2}
	\end{array} 
	\right.
\end{equation}
where
\begin{align}
	R_2^* &= 1 + \frac{\varepsilon}{2} + \ln (\frac{\varepsilon}{2}), \\
	R_H^* &= R_2^* - \int_{1}^{1 + \frac{\varepsilon}{2}}	\frac{1}{h_+(u)}du
	\\
	R_1^* &= R_H^* - \int_{1 - \frac{\varepsilon}{2}}^1	\frac{1}{h_{-}(u)}du .
\end{align}
Note that ${\rm atanh} (1-\frac{\varepsilon}{2}) \simeq \frac{1}{2}\ln(4/\varepsilon)$. In the calculation of the integrals (\ref{eq:Bohr_Sommerfeld_R}), (\ref{eq:Bohr_Sommerfeld_I}) we did not use the above expressions, but we substituted simply $dr^*=(1/h(r))dr$.

\section{Generalized Bohr-Sommerfeld Rule}\label{app:bohr-sommerfeld}

The bound states, namely the normal modes $E_n$, in a potential well $V(x)$ may be well approximated by the well-known method of the Bohr-Sommerfeld rule 
\begin{equation}\label{eq:BS_rule}
	\int_{x_0}^{x_1}\sqrt{E_{n} - V(x)} dx = \pi \left(n+\frac{1}{2} \right),
\end{equation}
where $x_0$, $x_1$ are the roots of the integrand, depending on $E_n$. This method can be derived from WKB theory for the Schr\"odinger equation. It has been used in the calculation of the high-overtone normal modes of the Schwarzschild black hole \cite{1990CQGra...7L..47G}.

In the case of quasi-stationary states in a partially confining potential like the one of equation (\ref{eq:V}), the Bohr-Sommerfeld rule (\ref{eq:BS_rule}) has been generalized in Ref \cite{1991PhLA..157..185P} as
\begin{equation}
	\int_{x_0}^{x_1}\sqrt{E_{n} - V(x)} dx = \pi \left(n+\frac{1}{2} \right) - \chi (w), 
\end{equation}
where 
\begin{equation}
	\chi (w) = \frac{1}{2} w(1-\ln (w)) + \frac{1}{4i} \ln\left( \frac{\Gamma (1/2+iw)}{\Gamma (1/2-iw) (1+e^{-2\pi w})}\right), 
	\quad
	w =\frac{1}{\pi} \int_{x_1}^{x_2} \sqrt{V(x) - E_n} dx,
\end{equation}
where $\Gamma$ denotes the Gamma function. Here $x_0$, $x_1$ denote the roots of $E_n - V(x) = 0$ that define the bounding region $E_n \geq V(x)$ and $x_2$ is the upper limit of the  reflecting region $E_n \leq V(x)$, as in Figure \ref{fig:V}. This expression can be simplified further for sufficiently low modes
 \cite{karnakov2013wkb} as
 \begin{equation}\label{eq:gen_BS_rule}
 	\int_{x_0}^{x_1}\sqrt{E_{n} - V(x)} dx = \pi \left(n+\frac{1}{2} \right) - \frac{i}{4}e^{-2\int_{x_1}^{x_2} \sqrt{V(x)-E_n} dx}.
 \end{equation}
The imaginary part is a measure of the barrier penetrability, that is absent in the normal Bohr-Sommerfeld rule (\ref{eq:BS_rule}) since the barrier is infinite in the latter case. 

We shall denote
\begin{equation}
	E_n = E_{ {\rm R},n} + i E_{ {\rm I},n}.
\end{equation}
In case the imaginary part is negligible with respect to the real part $E_{{\rm I},n} \ll E_{{\rm R},n}$ the generalized Bohr-Sommerfeld rule (\ref{eq:gen_BS_rule}) may be decomposed as follows
\begin{align}
	\label{eq:gen_BS_rule_R}
	&\int_{x_0}^{x_1} \sqrt{E_{{\rm R}, n} - V(x)} dx = \pi \left( n + \frac	{1}{2}\right),
	\\
	\label{eq:gen_BS_rule_I}
	&E_{{\rm I}, n} = - \frac{1}{2} \exp\left(-2\int_{x_1}^{x_2} \sqrt{V(x) - E_{{\rm R}, n}} dx\right)
	\left(\int_{x_0}^{x_1} \frac{1}{\sqrt{E_{{\rm R}, n} -  V(x)}} dx \right)^{-1},
\end{align}
which correspond to (\ref{eq:Bohr_Sommerfeld_R}), (\ref{eq:Bohr_Sommerfeld_I}). We used this generalized Bohr-Sommerfeld formulas to solve the Sturm-Liouville problem (\ref{eq:SL}). The accuracy of the method in the calculation of the quasi-normal modes of gravastars has been verified in Ref. \cite{2017CQGra..34l5006V}.

We calculated the integrals (\ref{eq:Bohr_Sommerfeld_R}), (\ref{eq:Bohr_Sommerfeld_I}) by a combination of analytical and numerical techniques. For $\varepsilon \ll 1$ the point $r_0$ lies within the de Sitter core $r_0/r_{\rm H} < 1 - \varepsilon/2$ and the points $r_1<r_2$ lie outside the black hole within the Schwarzschild spacetime $r_{1}/r_{\rm H} > 1 + \varepsilon/2$. In particular, we have
\begin{equation}
	u_0 = \frac{L}{E_{\rm R} + L}, \quad \text{where} \;
	L = \ell (\ell + 1)
\end{equation}
and we use the dimensionless variable $u=r/r_{\rm H}$. It is
\begin{equation}
	\int_{u_0}^{1-\frac{\varepsilon}{2}} \frac{1}{1-u^2}\sqrt{E_{\rm R} - L\frac{1-u^2}{u^2}} du = 
	\left[ \sqrt{E_{\rm R}}\, {\rm atanh}\left( \frac{u^2 - u_0^2}{1-u_0^2} \right)^{\frac{1}{2}} 
	- \sqrt{L}\, {\rm atan}\left( \frac{u^2}{u_0^2} - 1 \right)^{\frac{1}{2}} \right]_{u = 1-\frac{\varepsilon}{2}}.
\end{equation}
This diverges logarithmically for $\varepsilon \ll 1$ and therefore can be calculated analytically for a Compton or Planck or smaller $\varepsilon$ (see (\ref{eq:Compton}), (\ref{eq:Planck}))
\begin{equation}
	{\rm atanh}\left( \frac{u^2 - u_0^2}{1-u_0^2} \right)^{\frac{1}{2}} \simeq \frac{1}{2} \ln \frac{4(1-u_0^2)}{\varepsilon}.
\end{equation}
In the interval $r_{\rm H}-\alpha/ 2 \leq r \leq r_{\rm H} + \alpha/2$ it applies the density profile (\ref{eq:solutions}) and we use the variable $x=(r-r_{\rm H})/\alpha = (u-1)/\varepsilon = \mathscr{O}(1)$. The metric function $h_{\pm}$ is of order $\varepsilon$, $h_{\pm} = \mathscr{O}(\varepsilon)$. Thus we have
\begin{equation}\label{eq:int_pm}
	\int_{r_{\rm H} - \frac{\alpha}{2}}^{r_{\rm H} + \frac{\alpha}{2}} \frac{1}{h_{\pm}} \sqrt{E_{\rm R} - V(r)} dr = 
	\sqrt{E_{\rm R}}
	\int_{-\frac{1}{2}}^{+\frac{1}{2}} \frac{1}{h_{\pm}/\varepsilon} dx  + \mathscr{O}(\varepsilon).
\end{equation}
The function $h_{\pm}$ is Taylor expanded about $ \varepsilon=0$ and the integral is calculated numerically. We find numerically that for $K=1$ and $K=2$ all quasi-normal modes are constrained (as in Figure \ref{fig:omega_n=0_l=2}) by the two limiting solutions of the $K=1$, $N=3$ solutions that correspond to $A^{(\pm)}_3 = 8$, $A^{(\pm)}_3 = -4$ (see Appendix \ref{app:DE-BH}), which we call below I and II.
We have
\begin{align}
	\label{eq:rho_I}
	\rho_{(\pm)}^{\rm I} &= 8 x^3 \mp 6 x^2 + \frac{1}{2} + \mathscr{O}(\varepsilon),
	\\
	h_{(\pm)}^{\rm I} &= \varepsilon \left(- 6 x^4 \pm 6 x^3 - \frac{1}{2} x + \frac{3}{8} \right) + \mathscr{O}(\varepsilon^2)
\end{align}
and 
\begin{align}
	\label{eq:rho_II}
	\rho_{(\pm)}^{\rm II} &= - 4 x^3 \pm 6 x^2 - 3 x + \frac{1}{2} + \mathscr{O}(\varepsilon),
	\\
h_{(\pm)}^{\rm II} &= \varepsilon \left(3 x^4 \mp 6 x^3 + \frac{9}{2} x^2 - \frac{1}{2} x + \frac{3}{16} \right) + \mathscr{O}(\varepsilon^2)
\end{align}
and the integral (\ref{eq:int_pm}) is calculated numerically for both solutions I, II, decomposing it to regions $-1/2\leq x\leq 0$ where $h_{-}$ applies and $0\leq x\leq 1/2$ where $h_{+}$ applies.

There remains the region $1/2 \leq x \leq x_1$ corresponding to $r_{\rm H} + \alpha/2 \leq r \leq r_1 $, equivalently $1+\varepsilon /2 \leq u \leq u_1$. The radius $r_1$ is the smallest root within $r\geq r_{\rm H} + \alpha/2$ of
\begin{equation}
	\left( 1-\frac{r_{\rm H}}{r} \right) \left(L\frac{r_{\rm H}^2}{r^2} - 3\frac{r_{\rm H}^3}{r^3} \right) = 0.
\end{equation}
We write the integral
\begin{equation}
	I_{\rm Schw} = \int_{1+\frac{\varepsilon}{2}}^{u_1}\frac{1}{(1-\frac{1}{u})}\sqrt{E_{\rm R} - \left(1-\frac{1}{u} \right) \left(\frac{L}{u^2} - \frac{3}{u^3} \right) } du
\end{equation}
as
\begin{equation}
	I_{\rm Schw} = \int_{1+\frac{\varepsilon}{2}}^{u_1} \frac{du}{u(u-1)}\sqrt{P(u)},\quad
	\text{where}\;
	P(u) = E_{\rm R} u^4 - L u^2 + (L+3)u - 3  ,
\end{equation}
where $P(u_1)=0$. Considering that $(\ln (u-1))^\prime = 1/(u-1)$ we get
\begin{equation}
	I_{\rm Schw} = -\ln\left(\frac{\varepsilon}{2}\right) \cdot\frac{ \sqrt{ P\left(  1+\frac{\varepsilon}{2} \right)} }{1+\frac{\varepsilon}{2}}
	- \int_{1+\frac{\varepsilon}{2}}^{u_1} du\ln (u-1)\frac{d}{du}\frac{\sqrt{P(u)}}{u}.
\end{equation}
This expression is calculated numerically in a straightforward manner.

The integral in the numerator of equation (\ref{eq:Bohr_Sommerfeld_I}) is calculated directly numerically. The integral in the denominator of equation (\ref{eq:Bohr_Sommerfeld_I}) is calculated by using analogous treatments as above, except in the Schwarzschild region where there appears the additional pole at $u_1$. We approximated this non-regular integral 
\begin{equation}
	I_{{\rm Schw},2} = \int_{1+ \frac{\varepsilon}{2}}^{u_1}
	du\, \frac{u^3}{(u-1)\sqrt{P(u)}},		
\end{equation} 
by use of the transformation $u=y+1$ and expanding $P(y+1)$ with respect to $y$. We get
\begin{equation}
	I_{{\rm Schw},2} \simeq \frac{1}{\sqrt{E_{\rm R}}} \int_{\frac{\varepsilon}{2}}^{y_1}dy\, \frac{(y+1)^3}{y\sqrt{1-\frac{y}{y_1}}}
	= \frac{1}{\sqrt{E_{\rm R}}}\left( 2y_1^2 + 3y_1 + \ln\frac{8y_1}{\varepsilon}\right) + \mathscr{O}(\varepsilon) 
	,
	\;
	y_1 = \frac{E_{\rm R}}{L-3-4E_{\rm R}} .
\end{equation}

\end{document}